\documentstyle[12pt,epsfig]{article}
\begin{document}
\def\beq{\begin{equation}}
\def\eeq{\end{equation}}
\def\bea{\begin{eqnarray}}
\def\eea{\end{eqnarray}}
\def\ve{\vert}
\def\vel{\left|}
\def\ver{\right|}
\def\nnb{\nonumber}
\def\ga{\left(}
\def\dr{\right)}
\def\aga{\left\{}
\def\adr{\right\}}
\def\rar{\rightarrow}
\def\nnb{\nonumber}
\def\la{\langle}
\def\ra{\rangle}
\def\lla{\left<}
\def\rra{\right>}
\def\ba{\begin{array}}
\def\ea{\end{array}}
\def\tep{$B \rar K \ell^+ \ell^-$}
\def\tepm{$B \rar K \mu^+ \mu^-$}
\def\tept{$B \rar K \tau^+ \tau^-$}
\def\ds{\displaystyle}

\baselineskip 20pt        

\noindent
\hspace*{12cm}
{\bf (hep-ph/9910501)}\\
\noindent
\hspace*{12cm}
YUMS 99-024\\  
%\noindent
%\hspace*{12cm}
%METU-PHYS-HEP-99-xx\\ 

\vspace*{1.5cm}

\begin{center}
{\Large \bf A Systematic Analysis of the Exclusive 
           $B \rar K^\ast \ell^+ \ell^-$ Decay}\\

\vspace*{1cm}

{\bf T. M. Aliev $^{a,}$\footnote{taliev@metu.edu.tr}},~~ 
{\bf C. S. Kim $^{b,}$\footnote{kim@cskim.yonsei.ac.kr,~ 
http://phya.yonsei.ac.kr/\~{}cskim/}}~~and~~
{\bf Y. G. Kim$^{b,}$\footnote{ygkim@cskim.yonsei.ac.kr}}\\

\vspace*{0.5cm}
 $a:$ Physics Department, Middle East Technical University, 06531 Ankara, Turkey\\
 $b:$ Physics Department, Yonsei University, Seoul 120-749,
Korea~~~~~~~~~~~~~~~~~~~~~~~\\

\vspace{1.0cm}
 
(\today)
\vspace*{1.0cm}

\end{center}        

\begin{abstract}
\vspace{0.5cm}

\noindent 
A model--independent analysis for the exclusive, rare
$B \rar K^\ast \ell^+ \ell^-$ decay is presented. 
Systematically studied are  the experimentally measured quantities, 
such as, branching ratio, forward--backward asymmetry, 
longitudinal polarization of the final leptons, and
the ratio $\Gamma_L/\Gamma_T$ of the decay widths when $K^\ast$ meson is 
longitudinally and transversally polarized. 
The dependence of the asymmetry parameter 
$\alpha = 2 \Gamma_L/\Gamma_T -1$ on the new Wilson coefficients 
is also studied in detail.
It is found that the afore--mentioned physical
observables are quite sensitive to the new Wilson coefficients.
Therefore, once we have the experimental data with high statistics and a
deviation from the Standard Model, we can interpret the source of such
discrepancy.

\end{abstract}

\vspace{1cm}
%PACS numbers: 13.20.He, 11.55.Hx, 12.38.Cy

\newpage

\section{Introduction}

Experimental discovery of the inclusive and exclusive $B \rar X_s \gamma$
and $B \rar K^\ast \gamma$ decays \cite{R1} stimulated the study of the rare 
$B$ decays in a new manner. These decays take place via flavor--changing
neutral current (FCNC) transition of $b \rar s$, which occur only through loops
in the Standard Model (SM). For this reason the study of the FCNC decays can
provide sensitive test for investigation of the gauge structure of the SM at
loop level. At the same time these decays constitute quite a suitable tool in 
looking for new physics beyond the SM. New physics can appear in rare decays 
through the Wilson coefficients which can take values distinctly different from
their SM counterparts or through the new structure in effective Hamiltonian
(see for example Refs. \cite{R2}--\cite{R11}).

Currently the main interest is focused on the rare meson decays for which
the SM predicts ``large" branching ratios, and which can be potentially
measurable in the near future. The rare $B \rar K^\ast \ell^+ \ell^-~ 
(\ell=e,\mu,\tau)$ decays are such decays. For these decays the experimental
situation is very promising \cite{R12} with $e^+ e^-$ and hadron
colliders focusing only on the observation of exclusive modes with
$\ell=e,\mu$ and $\tau$ as the final states. At the quark level, the decay 
$B \rar K^\ast \ell^+ \ell^-$ is described by $b \rar s \ell^+ \ell^-$
transition. The inclusive $b \rar s \ell^+ \ell^-$ transition in framework 
of the specific extended models were investigated in many papers (see for
example \cite{R5,R11,R13,R14}). Note that the most general model independent
analysis of the $b \rar s \ell^+ \ell^-$ decay, in terms of 10 types of local 
four--Fermi interactions, was performed in Ref. \cite{R9}, 
which has been extended 
to include two more non-local interactions in Ref. \cite{R10}. 
New physics effects in the exclusive rare decays, 
$B \rar K^{(*)} \nu \nu$, have been systematically analyzed also 
in Ref. \cite{RR15}.

It is well known that theoretical analysis of the inclusive decays is easy
but their experimental detection is difficult. 
For exclusive decays the situation is reversed,  
{\it i.e.}, these decays can easily be studied in
experiments, but theoretically they have drawbacks and predictions are model
dependent. This is due to the fact that in calculating the branching 
ratios and other
observables for exclusive decays, we face the problem of computing the
matrix element of the effective Hamiltonian responsible for exclusive
decays, between initial and final hadron states. This problem is related to
the non--perturbative sector of QCD and can be solved only by means of a
non--perturbative approach. These matrix elements have been investigated 
in framework of different approaches such as chiral theory \cite{R15}, three
point QCD sum rules \cite{R16}, relativistic model by using light--front
formalism \cite{R17}, effective heavy quark theory \cite{R18} and light cone
QCD sum rules \cite{R19,R20}. 

The present paper is organized as follows: In Section 2 we give the most
general form of the effective Hamiltonian. 
Then, using this Hamiltonian and the helicity amplitude formalism, 
we calculate the differential decay width, including the lepton mass effects.
In this Section we also present the expressions of the other physical
observables, such as forward--backward asymmetry, and the ratio of the decay
widths when $K^\ast$ meson is polarized longitudinally and transversally.
Section 3 is devoted to the numerical analysis, and concluding remarks are 
also in Section 3.

\section{Theoretical Background}

The matrix element of the  $B \rar K^\ast \ell^+ \ell^-$ decay at the quark
level is described by $b \rar s \ell^+ \ell^-$ transition.
Following the work \cite{R9,R10}, we write the matrix element of the 
$b \rar s \ell^+ \ell^-$ transition as a sum of the SM and new physics
contributions, 
\bea
{\cal M} &=& {\cal M}_{\rm SM} + {\cal M}_{\rm new}~,
\eea
where ${\cal M}_{\rm SM}$ is the SM part and is given by
\bea
{\cal M}_{\rm SM} &=& \frac{G \alpha}{\sqrt{2} \pi} V_{tb}V_{ts}^\ast 
\Bigg\{ \ga C_9^{eff} - C_{10} \dr \bar s_L \gamma_\mu b_L \,
\bar \ell_L \gamma^\mu \ell_L +
\ga C_9^{eff} + C_{10} \dr \bar s_L \gamma_\mu b_L \,
\bar \ell_R \gamma^\mu \ell_R \nnb \\
&-& 2 C_7^{eff} \bar s i \sigma_{\mu\nu}  \frac{q^\nu}{q^2} 
\ga m_s L + m_b R \dr b \, \bar \ell \gamma^\mu \ell  \bigg\}~,
\eea
where $R=(1+\gamma_5)/2$ and $L=(1-\gamma_5)/2$, and all of the 
Wilson coefficients are evaluated at the scale $\mu = m_b = 4.8$~GeV. 

In Ref. \cite{R9}, it has been shown that 
there are ten independent local four--Fermi
interactions which may contribute to the process, and the explicit form of
${\cal M}_{\rm new}$ can be written as
\bea
\lefteqn{
{\cal M}_{\rm new} = \frac{G\alpha}{\sqrt{2} \pi}
 V_{tb}V_{ts}^\ast
\Bigg\{ C_{LL} \bar s_L \gamma_\mu b_L \, \bar \ell_L
\gamma^\mu \ell_L +
C_{LR} \bar s_L \gamma_\mu b_L \, \bar \ell_R \gamma^\mu \ell_R +
C_{RL} \bar s_R \gamma_\mu b_R \, \bar \ell_L \gamma^\mu \ell_L} \nnb \\
&&+ C_{RR} \bar s_R \gamma_\mu b_R \, \bar \ell_R \gamma^\mu \ell_R +
C_{LRLR} \bar s_L b_R \, \bar \ell_L  \ell_R 
+ C_{RLLR} \bar s_R b_L \, \bar \ell_L  \ell_R +     
C_{LRRL} \bar s_L b_R \, \bar \ell_R  \ell_L \nnb \\
&&+ C_{RLRL} \bar s_R b_L \, \bar \ell_R  \ell_L 
+ C_T \bar s \sigma_{\mu\nu} b \, \bar \ell \sigma^{\mu\nu} \ell +
i C_{TE} \bar s \sigma_{\mu\nu} b \, \bar \ell \sigma_{\alpha\beta} \ell
\epsilon^{\mu\nu\alpha\beta} \Bigg\}
\eea
It should be noted that in the present  analysis we will neglect the tensor
type interactions ({\it i.e.}, 
terms with coefficients $C_T$ and $C_{TE}$) since
the numerical analysis which is carried in
Ref. \cite{R9} shows that the physical observables are not sensitive to
the presence of the tensor interactions. 

From Eq. (1), in order to calculate the decay width for the
exclusive $B \rar K^\ast \ell^+ \ell^-$ decay, the following matrix elements 
\bea
&&\lla K^\ast \vel \bar s \gamma_\mu (1 \pm \gamma_5) b \ver B \rra, \nnb \\
&&\lla K^\ast \vel \bar i s \sigma_{\mu\nu} q^\nu (1 + \gamma_5) b \ver B \rra,
~~~{\rm (strange~ quark~ mass~ is~ neglected)} \nnb \\
{\rm and} ~~~
&&\lla K^\ast \vel \bar s (1 \pm \gamma_5) b \ver B \rra \nnb
\eea
have to be calculated. 
These matrix elements can be written in terms of the form
factors in the following way
\bea
\lefteqn{
\lla K^\ast(p_{K^\ast},\varepsilon) \vel \bar s \gamma_\mu (1 \pm \gamma_5) b \ver
B(p_B) \rra =} \nnb \\
&&- \epsilon_{\mu\nu\rho\sigma} \varepsilon^{\ast\nu} p_{K^\ast}^\rho q^\sigma 
\frac{2 V(q^2)}{m_B+m_{K^\ast}} \pm i \varepsilon_\mu^\ast (m_B+m_{K^\ast})
A_1(q^2) \mp i (p_B + p_{K^\ast})_\mu (\varepsilon^\ast q)
\frac{A_2(q^2)}{m_B+m_{K^\ast}} \nnb \\
&&\mp i q_\mu \frac{2 m_{K^\ast}}{q^2} (\varepsilon^\ast q) 
\left[A_3(q^2)-A_0(q^2)\right]~,  \\ 
\lefteqn{
\lla K^\ast(p_{K^\ast},\varepsilon) \vel \bar s i \sigma_{\mu\nu} q^\nu (1 + \gamma_5) b \ver
B(p_B) \rra =} \nnb \\
&& 4 \epsilon_{\mu\nu\rho\sigma} \varepsilon^{\ast\nu} p_{K^\ast}^\rho q^\sigma 
T_1(q^2) + 2 i \left[ \varepsilon_\mu^\ast (m_B^2-m_{K^\ast}^2) -
(p_B + p_{K^\ast})_\mu (\varepsilon^\ast q) \right] T_2(q^2) \nnb \\ 
&+& 2 i (\varepsilon^\ast q) \left[ q_\mu -
(p_B + p_{K^\ast})_\mu \frac{q^2}{m_B^2-m_{K^\ast}^2} \right] T_3(q^2)~,
\eea
where $\varepsilon$ is the polarization vector of $K^\ast$ meson, and 
$q=p_B-p_{K^\ast}$ is the momentum transfer.
In order to ensure finiteness of (4) at $q^2=0$, we demand that 
$A_3(q^2=0) = A_0(q^2=0)$.
For calculation of the matrix element 
$\lla K^\ast \vel \bar s (1 \pm \gamma_5) b \ver B \rra$, we
multiply both sides of Eq. (4) by $q_\mu$ and use equation of motion.
Neglecting the strange quark mass, we get
\bea
\lefteqn{
\lla K^\ast(p_{K^\ast},\varepsilon) \vel \bar s (1 \pm \gamma_5) b \ver
B(p_B) \rra =} \nnb \\ 
&&\frac{1}{m_b} \Big\{ \mp i (\varepsilon^\ast q) (m_B+m_{K^\ast}) A_1(q^2)
\pm i (m_B-m_{K^\ast}) (\varepsilon^\ast q) A_2(q^2) \nnb \\
&&\pm 2 i m_{K^\ast} (\varepsilon^\ast q) \left[A_3(q^2)-A_0(q^2)\right]~\Big\}.
\eea
Using the equation of motion, the form factor $A_3$ can be written as a
linear combination of the form factors $A_1(q^2)$ and $A_2(q^2)$ 
(see Ref. \cite{R16})
\bea
A_3(q^2) = \frac{m_B+m_{K^\ast}}{2 m_{K^\ast}} A_1(q^2) - 
\frac{m_B-m_{K^\ast}}{2 m_{K^\ast}} A_2(q^2)~. \nnb 
\eea
Substituting this relation in the matrix element 
$\lla K^\ast \vel \bar s (1 \pm \gamma_5) b \ver B \rra$, we get 
\bea
\lla K^\ast(p_{K^\ast},\varepsilon) \vel \bar s (1 \pm \gamma_5) b \ver B(p_B) \rra
= \frac{1}{m_b} \Big\{\mp 2 i m_{K^\ast} (\varepsilon^\ast q) A_0(q^2)  
\Big\}~.
\eea
Finally,  
for the matrix elements of $B \rar K^\ast \ell^+ \ell^-$ decay we have
\bea
\lefteqn{
{\cal M} = \frac{G \alpha}{4 \sqrt{2} \pi} V_{tb} V_{ts}^\ast \Bigg\{
\Bigg[-\epsilon_{\mu\nu\rho\sigma} \varepsilon^{\ast\nu} p_{K^\ast}^\rho q^\sigma 
\frac{2 V(q^2)}{m_B+m_{K^\ast}}-
i \varepsilon_\mu^\ast (m_B+m_{K^\ast}) A_1(q^2)} \nnb \\
&&+ i (p_B+p_{K^\ast})_\mu (\varepsilon^\ast q)
\frac{A_2(q^2)}{m_B+m_{K^\ast}}
+ i q_\mu \frac{2 m_{K^\ast}}{q^2} (\varepsilon^\ast q)(A_3 - A_0) \Bigg] \nnb \\
&&~~\times \Bigg[ (C_9^{eff}-C_{10}+C_{LL}) \bar \ell \gamma_\mu
(1-\gamma_5) \ell + (C_9^{eff}+C_{10}+C_{LR}) \bar \ell \gamma_\mu
(1+\gamma_5) \ell \bigg] \nnb \\
&&- 4 \frac{C_7^{eff}}{q^2} m_b \Bigg[ 4 \epsilon_{\mu\nu\rho\sigma} 
\varepsilon^{\ast\nu} p_{K^\ast}^\rho q^\sigma T_1(q^2) + 2 i \Big(\varepsilon_\mu^\ast
(m_B^2-m_{K^\ast}^2) + (p_B+p_{K^\ast})_\mu (\varepsilon^\ast q) \Big)
T_2(q^2) \nnb \\
&&~~~~~~~~~+2 i (\varepsilon^\ast q) \Bigg( q_\mu - (p_B+p_{K^\ast})_\mu
\frac{q^2}{m_B^2-m_{K^\ast}^2} \Bigg) T_3(q^2) \Bigg] \bar \ell \gamma_\mu\ell \\
&&+\Bigg[-\epsilon_{\mu\nu\rho\sigma} \varepsilon^{\ast\nu} p_{K^\ast}^\rho
q^\sigma \frac{2 V(q^2)}{m_B+m_{K^\ast}}
 + i \varepsilon_\mu^\ast (m_B+m_{K^\ast}) A_1(q^2) - i (p_B+p_{K^\ast})_\mu
(\varepsilon^\ast q) \frac{A_2(q^2)}{m_B+m_{K^\ast}} \nnb \\
&&~~~~~~~~~-i q_\mu \frac{2 m_{K^\ast}}{q^2} (\varepsilon^\ast q) 
(A_3(q^2) - A_0(q^2)) \Bigg] \Big[ C_{RL} \bar \ell \gamma_\mu(1-\gamma_5) \ell 
+ C_{RR} \bar \ell \gamma_\mu (1+\gamma_5) \ell \Big] \nnb \\
&&+\frac{1}{m_b} \Big[-2 i m_{K^\ast} (\varepsilon^\ast q) 
A_0(q^2) \Big] \Big[ (C_{LRLR} - C_{RLLR})  \bar \ell (1+\gamma_5) \ell +
(C_{LRRL} - C_{RLRL})  \bar \ell (1-\gamma_5) \ell \Big] \Bigg\}~. \nnb 
\eea  

Using the matrix element of $B \rar K^\ast \ell^+ \ell^-$ decay 
(see Eq. (8)) and  
the helicity amplitude formalism (for more detail see Refs.
\cite{R21,R22}) for the differential decay rate width, we get
\bea
\lefteqn{
\frac{d\Gamma}{dq^2 dx} = \frac{G^2 \alpha^2}{2^{14} \pi^5 m_B^3} 
\vel V_{tb} V_{ts}^\ast \ver^2
v \lambda^{1/2}(m_B^2,q^2,m_{K^\ast}^2)} \nnb \\
&&\times \Bigg\{ 
\vel {\cal M}^{+-}_{+}\ver^2 +
\vel {\cal M}^{+-}_{-}\ver^2 +
\vel {\cal M}^{++}_{-}\ver^2 +
\vel {\cal M}^{++}_{+}\ver^2 +
\vel {\cal M}^{-+}_{+}\ver^2 +
\vel {\cal M}^{-+}_{-}\ver^2 +
\vel {\cal M}^{--}_{+}\ver^2 \nnb \\ 
&&+
\vel {\cal M}^{--}_{-}\ver^2 +
\vel {\cal M}^{++}_{0}\ver^2 +
\vel {\cal M}^{+-}_{0}\ver^2 +
\vel {\cal M}^{-+}_{0}\ver^2 +
\vel {\cal M}^{--}_{0}\ver^2 \Bigg\}~,
\eea
where superscripts denote helicities of the leptons and subscripts
correspond to the helicity of the $K^\ast$ meson.
In Eq. (9),
\bea 
&&\lambda(m_B^2,q^2,m_{K^\ast}^2) = m_B^4 + m_{K^\ast}^4 +
q^4 - 2 m_B^2 q^2 - 2 m_B^2 m_{K^\ast}^2 - 2 m_{K^\ast}^2 q^2~, \nnb \\ 
&&q^2=(p_B-p_{K^\ast})^2~, \nnb \\ 
&&v = \sqrt{1 - 4 m_\ell^2/q^2},~~{\rm (velocity~ of~ the~ lepton),~ and} \nnb \\
&&x=\cos\theta,~~  (\theta = {\rm angle~ between}~ K^\ast~{\rm and}~ \ell^-). \nnb
\eea 
The explicit forms of 
${\cal M}^{\lambda_\ell \,  \lambda_\ell}_{\lambda_V}$ are as follows:
\bea
{\cal M}^{++}_{\pm} &=& \pm \sqrt{2} m_\ell \sin\theta \Big\{
( 2 C_9^{eff} + C_{LL} + C_{LR} ) H_\pm +
4 C_7^{eff} \frac{m_b}{q^2} {\cal H}_\pm + (C_{RR} + C_{RL}) h_\pm\Big\}~,
\\ \nnb \\
{\cal M}^{+-}_{\pm} &=& (-1\pm\cos\theta) \sqrt{\frac{q^2}{2}} \Big\{
\Big[ 2 C_9^{eff} + C_{LL} + C_{LR}  + v (2 C_{10} + C_{LR} - C_{LL} )
\Big] H_\pm\nnb \\
&&+ 4 C_7^{eff} \frac{m_b}{q^2} {\cal H}_\pm +
\Big[ C_{RL} + C_{RR} + v ( C_{RR} - C_{RL} ) \Big] h_\pm \Big\}~,
\\ \nnb \\
{\cal M}^{-+}_{\pm} &=& (1\pm\cos\theta) \sqrt{\frac{q^2}{2}} \Big\{
\Big[ 2 C_9^{eff} + C_{LL} + C_{LR}  + v (- 2 C_{10} + C_{LL} - C_{LR} )        
\Big] H_\pm\nnb \\
&&+ 4 C_7^{eff} \frac{m_b}{q^2} {\cal H}_\pm + 
\Big[ C_{RL} + C_{RR} + v ( C_{RL} - C_{RR} ) \Big] h_\pm \Big\}~,
\\\nnb \\
{\cal M}^{--}_{\pm} &=& (\mp \sqrt{2} m_\ell \sin\theta) \Big\{
( 2 C_9^{eff} + C_{LL} + C_{LR} ) H_\pm +           
4 C_7^{eff} \frac{m_b}{q^2} {\cal H}_\pm \nnb \\
&&+ (C_{RL} + C_{RR}) h_\pm\Big\}~,
\\ \nnb \\
{\cal M}^{++}_{0} &=& 2 m_\ell \cos\theta \Big\{
( 2 C_9^{eff} + C_{LL} + C_{LR} ) H_0 -           
4 C_7^{eff} \frac{m_b}{q^2} {\cal H}_0 + (C_{RL} + C_{RR}) h_0\Big\} \nnb \\
&&+ 2 m_\ell \Big\{ (2 C_{10} - C_{LL} + C_{LR} ) H_S^0 
+(-C_{RL} + C_{RR}) h_S^0 \Big\} \\
&&+\frac{2}{m_b} \sqrt{q^2} \Big\{ \Big[\sqrt{q^2} (1-v) (C_{LRLR} - C_{RLLR}) -
\sqrt{q^2} (1+v) (C_{LRRL} - C_{RLRL}) \Big] H_S^0 \Big\}~,
\nnb \\ \nnb \\
{\cal M}^{+-}_{0} &=& - \sqrt{q^2} \sin\theta \Big\{
\Big[ (C_9^{eff}- C_{10}+ C_{LL})(1-v) + (C_9^{eff}+ C_{10}+ C_{LR})
(1+v)\Big] H_0 \nnb \\ 
&&- 4 C_7^{eff} \frac{m_b}{q^2} {\cal H}_0 
+ \Big[C_{RL}(1-v) + C_{RR}(1+v)\Big] h_0 \Big\}~,
\\ \nnb \\
{\cal M}^{-+}_{0} &=&  - \sqrt{q^2} \sin\theta \Big\{
\Big[ (C_9^{eff}- C_{10}+ C_{LL})(1+v) + (C_9^{eff}+ C_{10}+ C_{LR})
(1-v)\Big] H_0 \nnb \\ 
&&- 4 C_7^{eff} \frac{m_b}{q^2} {\cal H}_0        
+ \Big[C_{RL}(1+v) + C_{RR}(1-v)\Big] h_0 \Big\}~,
\\ \nnb \\
{\cal M}^{--}_{0} &=& - 2 m_\ell \cos\theta \Big\{         
( 2 C_9^{eff} + C_{LL} + C_{LR} ) H_0 -             
4 C_7^{eff} \frac{m_b}{q^2} {\cal H}_0 + (C_{RL} + C_{LL}) h_0\Big\} \nnb \\
&&+ 2 m_\ell \Big\{ (2 C_{10} - C_{LL} + C_{LR} ) H_S^0 
+(C_{RR} - C_{RL}) h_S^0 \Big\} \\
&&+\frac{2}{m_b} \sqrt{q^2} \Big\{ \Big[\sqrt{q^2} (1+v) (C_{LRLR} -
C_{RLLR}) -
\sqrt{q^2} (1-v) (C_{LRRL} - C_{RLRL}) \Big] H_S^0 \Big\}~,\nnb 
\eea
where
\bea
H_\pm &=& \pm \lambda^{1/2} \frac{V(q^2)}{m_B+m_{K^\ast}} + 
(m_B+m_{K^\ast}) A_1(q^2)~, \\ \nnb \\
H_0 &=& \frac{1}{2 m_{K^\ast}\sqrt{q^2}} \Bigg[
- (m_B^2-m_{K^\ast}^2-q^2) (m_B+m_{K^\ast}) A_1(q^2) +
 \lambda\frac{A_2(q^2)}{m_B+m_{K^\ast}} \Bigg]~, \\ \nnb \\   
H_S^0 &=&\frac{ \lambda^{1/2}}{2 m_{K^\ast}\sqrt{q^2}}
\Bigg[-(m_B+m_{K^\ast}) A_1(q^2) + \frac{A_2(q^2)}{m_B+m_{K^\ast}} 
(m_B^2-m_{K^\ast}^2) \nnb \\
&+& 2 m_{K^\ast} \left[ A_3(q^2)-A_0(q^2)\right]~,  \nnb \\ 
&\equiv& \frac{ \lambda^{1/2}}{2 m_{K^\ast}\sqrt{q^2}}
\left[-2m_{K^\ast} A_0(q^2) \right] \\ \nnb \\  
{\cal H}_\pm &=& 2 \left[ \pm \lambda^{1/2} T_1(q^2) +
(m_B^2-m_{K^\ast}^2)T_2(q^2) \right]~, \\ \nnb \\ \nnb
{\cal H}_0 &=&   \frac{1}{m_{K^\ast}\sqrt{q^2}} 
\Bigg\{ (m_B^2-m_{K^\ast}^2) (m_B^2-m_{K^\ast}^2-q^2) T_2(q^2) \nnb \\
&-& \lambda \left[ T_2(q^2)
+ \frac{q^2}{m_B^2-m_{K^\ast}^2} T_3(q^2) \right] \Bigg\}~, \\ \nnb \\
h_\pm &=& H_\pm \left( A_1 \rar - A_1,~~A_2 \rar - A_2 \right)~,  \\
h_0 &=& H_0 \left( A_1 \rar - A_1,~~A_2 \rar - A_2 \right)~.
\eea

In the present paper, we study the dependence of the following measurable physical
quantities, such as \\
\hspace*{0.5cm} (i) $\Gamma_+/\Gamma_-$, \\
\hspace*{0.5cm} (ii) $\Gamma_L/\Gamma_T=\Gamma_0/(\Gamma_+ + \Gamma_-)$, \\
\hspace*{0.5cm} (iii)  the polarization parameter 
$\left[ 2 \Gamma_0/(\Gamma_+ + \Gamma_-) -1 \right]$, and \\
\hspace*{0.5cm} (iv) the lepton forward--backward asymmetry 
and the longitudinal lepton polarization,  \\
on the different ``new" Wilson coefficients. 
Here the subscripts in the decay width denotes the helicities
of the $K^\ast$ meson.
From Eq. (9), we can easily obtain the explicit
expressions for $\Gamma_+$, $\Gamma_-$ and $\Gamma_0$ as
\bea
\Gamma_\pm &=& \frac{G^2 \alpha^2}{2^{14}\pi^5 m_B^3}
\vel V_{tb}V_{ts}^\ast \ver^2 \int dq^2\int dx \, v \lambda^{1/2} \Bigg\{
\vel {\cal M}_\pm^{+-} \ver^2 + \vel {\cal M}_\pm^{++} \ver^2 \nnb \\
&+& \vel {\cal M}_\pm^{-+} \ver^2+\vel {\cal M}_\pm^{--} \ver^2 \Bigg\}~,
\eea
where the upper(lower) subscript in $\Gamma$ corresponds to 
${\cal M}_+({\cal M}_-)$ and
\bea
\Gamma_0 &=& \frac{G^2 \alpha^2}{2^{14}\pi^5 m_B^3}
\vel V_{tb}V_{ts}^\ast \ver^2 \int dq^2\int dx \, v \lambda^{1/2} \Bigg\{
\vel {\cal M}_0^{+-} \ver^2 + \vel {\cal M}_0^{++} \ver^2 \nnb \\
&+& \vel {\cal M}_0^{-+} \ver^2+\vel {\cal M}_0^{--} \ver^2 \Bigg\}~.
\eea
From Eqs. (25) and (26), the expressions for the ratios
$\Gamma_+/\Gamma_-$, $\Gamma_L/\Gamma_T=\Gamma_0/(\Gamma_+ + \Gamma_-)$
and the polarization parameter, which is equal to
$\alpha \equiv 2 \Gamma_L/\Gamma_T -1$, can easily be obtained. 
These quantities are separately measurable from the experiments.
In further analysis we will study the dependence of the branching ratio on
new Wilson coefficients which are related to the decay width by the relation
${\cal BR} (B \rar K^\ast \ell^+ \ell^-) = \Gamma(B \rar K^\ast \ell^+
\ell^-)~\tau_B$, where $\tau_B$ is the life time of the $B$ meson.  

The lepton forward--backward asymmetry, $A_{FB}$, is one of the most
useful tools in search of new physics beyond the SM. 
Especially the determination 
of the position of the zero value for $A_{FB}$ can predict possibly 
new physics contributions.
Indeed, existence of the new physics can be confirmed by the shift in the
position of the zero value of the forward--backward asymmetry \cite{R7}.  
Therefore, in the present work we analyze with special emphasis
the dependence of $A_{FB}$ on the different ``new" Wilson coefficients. 
The lepton forward--backward asymmetry is defined in the
following way
\bea
\frac{d}{dq^2}A_{FB}(q^2) = \frac{\displaystyle{
\int_0^1dx \frac{d\Gamma}{dq^2 dx} - \int_{-1}^0dx \frac{d\Gamma}{dq^2 dx}}}
{\displaystyle{
\int_0^1dx \frac{d\Gamma}{dq^2 dx} + \int_{-1}^0dx \frac{d\Gamma}{dq^2 dx}}}~.
\eea
Another very informative quantity in search of new physics is the final
lepton polarization, as shown in  Ref. \cite{R10}. 
Here we restrict ourselves only to the study 
of the longitudinal polarization of the $\tau$--lepton. The expression for
longitudinal polarization can be calculated from Eq. (9),
\bea
\lefteqn{P_L = }\nnb  
\eea
\bea
\frac{
\displaystyle{\int_0^1 dx \,\Bigg\{
\left[ \vel {\cal M}_\pm^{-+} \ver^2+\vel {\cal M}_\pm^{--} \ver^2 +
       \vel {\cal M}_0^{-+} \ver^2+\vel {\cal M}_0^{--} \ver^2 \right] -
\left[ \vel {\cal M}_\pm^{+-} \ver^2+\vel {\cal M}_\pm^{++} \ver^2 +
       \vel {\cal M}_0^{+-} \ver^2+\vel {\cal M}_0^{++} \ver^2 \right]
\Bigg\} v \lambda^{1/2}} } {
\displaystyle{\int_0^1 dx \,\Bigg\{
\left[ \vel {\cal M}_\pm^{-+} \ver^2+\vel {\cal M}_\pm^{--} \ver^2 +
       \vel {\cal M}_\pm^{+-} \ver^2+\vel {\cal M}_\pm^{++} \ver^2 +
       \vel {\cal M}_0^{-+} \ver^2+\vel {\cal M}_0^{--} \ver^2  +
       \vel {\cal M}_0^{+-} \ver^2+\vel {\cal M}_0^{++} \ver^2 \right]
\Bigg\} v \lambda^{1/2}} } ~. \nnb    
\eea

\section{Numerical Analysis and Conclusions}

Having the explicit expressions for the physically measurable quantities,
in this Section we will study the dependence of these quantities on the new
Wilson coefficients in ${\cal M}_{\rm new}$, Eq. (3). 
The values of the main input parameters, which appear in the expression for
the decay widths $\Gamma_0,~\Gamma_+,~\Gamma_-,~A_{FB}$ and the polarization
parameter $\alpha$, are: 
\bea
m_b &=& 4.8~{\rm GeV},~~~m_c=1.35~{\rm GeV},~~~m_\tau=1.78~{\rm GeV}, \nnb \\
m_\mu &=& 0.105~{\rm GeV},~~~m_B=5.28~{\rm GeV},~~~m_{K^\ast}=0.892~{\rm GeV}. \nnb
\eea 
We use the following
values for the Wilson coefficients of the SM:
$$
C_9^{\rm NDR}=4.153,~~~C_{10}=-4.546,~~~C_7=-0.311, 
$$
which correspond to the next-to-leading QCD corrections \cite{R23,R24}. 
The renormalization point $\mu$ and
the top quark mass are set to be 
$$
\mu=m_b=4.8~{\rm GeV},~~~m_t=175~{\rm GeV}. 
$$
(We follow Refs. \cite{R25}--\cite{R29} in taking into account 
the long--distance effects of the charmonium states). 
For the form factors, we have used the
results of the works \cite{R19,R20}.
Here we would like to stress
that, throughout numerical analysis the central values of the input parameters
are used and their theoretical errors, especially the ones related to the
form factors, might be sizeable, but are not taken into account in the
present work.     

Let us first study the change in the differential decay rate when the
corresponding Wilson coefficients change. We assume that all new Wilson
coefficients $C_{X}$ are real, {\it i.e.}, 
we do not introduce any new physics phase
in addition to the one present in the SM. 
In Figs. 1--3 (Figs. 4--6), we change $C_{LL},~C_{LR},~C_{RR},
~C_{RL},~C_{LRLR}$ and $C_{LRRL}$ for the $B \rar K^\ast \mu^+ \mu^-$
($B \rar K^\ast \tau^+ \tau^-$) decays. 
From these Figures, we can easily see that,
far from resonance regions, 
$d{\cal BR}/dq^2$ is more strongly dependent on $C_{LL}$ and
also on $C_{RL}$ than on the other $C_X$'s. 
This behavior can be explained as follows: \\
\hspace*{0.5cm}  
(i) Considering $B \rar K^\ast \mu^+ \mu^-$ decay, and 
neglecting the terms proportional to the lepton mass, the terms coming from 
$C_{LL}$ and $C_{RL}$ are (see Eqs. (10)--(17))
\bea
\vel {\cal M}_{C_{LL}} \ver^2 &=& (1\pm\cos\theta)^2 \frac{q^2}{2}
\vel2(C_9^{eff} -C_{10} +C_{LL})H_\pm  +
4 C_7^{eff} \frac{m_b}{q^2} {\cal H}_\pm \ver^2 \nnb \\
&+& \sin^2\theta~ q^2 \vel 2 (C_9^{eff} -C_{10} + C_{LL}) H_0 - 
4 C_7^{eff} \frac{m_b}{q^2} {\cal H}_0\ver^2~, \\ \nnb \\
\vel {\cal M}_{C_{RL}} \ver^2 &=& (1\pm\cos\theta)^2 \frac{q^2}{2}
\vel \Big[ 2 (C_9^{eff} -C_{10}) H_\pm + 4 C_7^{eff} \frac{m_b}{q^2} {\cal H}_\pm + 2
C_{RL} h_\pm \ver^2 \nnb \\
&+& \sin^2\theta~q^2 \vel 2 (C_9^{eff} -C_{10}) H_0  -
4 C_7^{eff} \frac{m_b}{q^2} {\cal H}_0 + 2 C_{RL} h_0  \ver^2~.
\eea
Far from the resonance region, for example $q^2 \simeq 5$~GeV$^2$,
Re$(C_9^{eff}-C_{10})\simeq 9.5$ and Re$(C_9^{eff}+C_{10})\simeq 0.4$.
Therefore, the interference terms between the terms proportional to 
$(C_9^{eff}-C_{10})$ and $C_{LL}$ ($C_{RL}$) are large and for
this reason the contributions coming from $C_{LL}$ and $C_{RL}$ are large. 
From these Figures we also see that the contribution of $C_{LL}$ is
constructive (destructive) when $C_{LL}= \vel C_{10} \ver~(C_{LL}= - \vel
C_{10} \ver)$. The situation for $C_{RL}$ is opposite to the previous case,
{\it i.e.}, its contribution is constructive (destructive) when $C_{LL}= - \vel
C_{10} \ver~(C_{LL}= \vel C_{10} \ver)$. \\
\hspace*{0.5cm} 
(ii) For the $B \rar K^\ast \tau^+ \tau^-$ decay the situation is similar to
the $B \rar K^\ast \mu^+ \mu^-$ transition, but slightly different.
Namely, in this case the largest contribution comes from $C_{LL}$ and the
contribution of the $C_{RL}$ becomes equal to the contributions that come
from $C_{RR},~C_{LR}$,  and {\it etc.} 
This situation can be explained by the fact the term
$\sim (1-v^2)$, which is very small for the muon case, gives destructive
contribution in the SM. 

In Fig. 7, we investigate the dependence of the partially
integrated branching ratio ${\cal BR}$ on the new Wilson coefficients. The
range for the integration is chosen $1$~GeV$^2 < q^2 < 8$~GeV$^2$ for the 
$B \rar K^\ast \mu^+ \mu^-$ decay and $15$~GeV$^2 < q^2 < 20$~GeV$^2$ for 
$B \rar K^\ast \tau^+ \tau^-$ channel, in order to avoid the
long distance contributions due to the $J/\psi$ and its excitations. 
For the $B \rar K^\ast \mu^+ \mu^-$ case, it follows from Fig. 7 that
the partially integrated branching ratio ${\cal BR}$ depends strongly 
on $C_{LL}$ and $C_{RL}$, but for the $B \rar K^\ast \tau^+ \tau^-$ decay 
it depends strongly only on $C_{LL}$, which is consistent with
the previous results for $d{\cal BR}/dq^2$. 
Dependence on the other coefficients is rather weak.
From these Figures it follows that the contributions of $C_{LL}$ and
$C_{RL}$ to ${\cal BR}$ are positive for $C_{LL}>0$ and $C_{RL}<0$, and
negative for $C_{LL}<0$ and $C_{RL}>0$. 

In Figs. 8--10 (Figs. 11--13)
we plot the dependence of the lepton forward--backward  asymmetry on the 
new Wilson
coefficients, within the range $-\vel C_{10}\ver \le C_{X} \le \vel C_{10}\ver $, 
for the  $B \rar K^\ast \mu^+ \mu^-$ ($B \rar K^\ast \tau^+ \tau^-$) decay.
The experimental bounds on the  branching ratio of the 
$B \rar K^\ast \mu^+ \mu^-$ and the  $B_s \rar \mu^+ \mu^-$
decays \cite{R30} suggest that this is the right order of magnitude range for
the vector and scalar Wilson coefficients.
For the $B \rar K^\ast \mu^+ \mu^-$ case, it follows form Figs. 8--10
that the lepton forward-backward asymmetry is more sensitive to the 
$C_{LL},~C_{LR}$ and $C_{RL}$ than to the other $C_{X}$'s. 
We emphasize that  when $C_{LL}$ and $C_{LR}$ are positive 
then the zero point of 
$dA_{FB}/dq^2$ is shifted to the right, and when $C_{LL}$ and $C_{LR}$
are negative, it shifts to the left  from its corresponding SM value. 
In other words, the determination of the zero point of the 
differential asymmetry tells us not only about the existence of new physics, 
but it also can
fix the sign of the new Wilson coefficients. From these Figures, we also see
that the lepton forward--backward asymmetry has a weak dependence on the
other Wilson coefficients. 
From Figs. 11--13, we can deduce the following results for the 
$B \rar K^\ast \tau^+ \tau^-$ decay: \\
\hspace*{0.5cm} 
(i) Position of the zero value of the $dA_{FB}/dq^2$ for the $B \rar K^\ast
\tau^+ \tau^-$ decay can be useful for extracting only $C_{LR}$. \\
\hspace*{0.5cm} 
(ii) The value of the $dA_{FB}/dq^2$ is very sensitive (excluding the
resonance region) to $C_{RR}$ and $C_{LRRL}$. In other words, analyzing the
zero point and magnitude of the $dA_{FB}/dq^2$ allows us in principle to
determine different $C_X$'s. 

As we have noted earlier, the experimentally measurable quantities, 
$\Gamma_+/\Gamma_-$, $\Gamma_L/\Gamma_T$ and $P_L$, can be useful for
distinguishing the effects of new physics from the ones of the SM. 
In Figs. 14--15, we present the dependence of the ratios
$\Gamma_+/\Gamma_-$ and $\Gamma_L/\Gamma_T$ on $C_X$'s for the 
$B \rar K^\ast \mu^+ \mu^-$ and $B \rar K^\ast \tau^+ \tau^-$ decays, 
respectively. The main difference compared to the previous analysis is that
the values $\Gamma_+/\Gamma_-$ and $\Gamma_L/\Gamma_T$ are more sensitive to the
coefficient $C_{RL}$. (The result for the SM can be obtained by substituting
$C_X=0$.)
From these Figures, we observe that the ratio $\Gamma_L/\Gamma_T$, when
$C_{RL}$ is varied between -4 and 4, changes between 1 and 4.5. Therefore,
the measurement of this ratio in experiments can yield unambiguous
information about the existence of new physics. In the 
$B \rar K^\ast \tau^+ \tau^-$ decay, the ratio $\Gamma_+/\Gamma_-$ is
again more sensitive to the coefficient $C_{RL}$, while the ratio 
$\Gamma_L/\Gamma_T$ is more sensitive to the coefficients $C_{LRLR}$ and 
$C_{LRRL}$.

Finally, in Fig. 16 we present the dependence of the longitudinal
polarization  $P_L$ of $\tau$ on the new coefficients $C_X$'s. 
We see that  $P_L$ is sensitive to all the coefficients except the
coefficient $C_{LRLR}$. 
The dependence of $P_L$ on different coefficients is not the same. For example,
$P_L$ always increases when $C_{RL}$ and $C_{LRRL}$ change in the region
$(-4,~4)$. However, $P_L$ first decreases when
$C_{LL},~C_{LR}$ and $C_{RR}$ increase from -4 to 0, and then  increases
when the coefficients increase from 0 to 4. 

%\section{Conclusion}

To summarize, 
in the present work the most general model independent analysis of the 
exclusive  $B \rar K^\ast \ell^+ \ell^-$ decay is presented. This
exclusive decay is known to be very clean experimentally and will be
measured at the present asymmetric $B$ factories 
and future hadronic $B$ factories, HERA-B, B-TeV and LHC-B. Moreover, the 
$B \rar K^\ast \ell^+ \ell^-$ decay is very sensitive to the various
extensions of the Standard Model. We have studied the 
$B \rar K^\ast \ell^+ \ell^-$ decay in a model independent manner. 
The sensitivity to the new coefficients of the differential and partially 
integrated branching ratios, and forward--backward asymmetries are
systematically studied. It is observed that the differential and partially
integrated branching ratio for  $B \rar K^\ast \mu^+ \mu^-$ decay is
more strongly dependent on $C_{LL}$ and $C_{RL}$ than on the other $C_{X}$'s. 
The reason for such a
strong dependence can be explained by the large interference between the
terms proportional to $(C_9^{eff}$--$C_{10})$ and $C_{LL}~(C_{RL})$.
For $B \rar K^\ast \tau^+ \tau^-$ case, the partially integrated
differential branching ratio is most sensitive to $C_{LL}$. This situation
can be explained by the fact that the terms $\sim (1-v^2)$ give destructive
contribution and, therefore, the contributions of the terms $\sim C_{RL}$
practically become equal to the contributions from the other coefficients. 
From an analysis of the position of the zero value of the lepton
forward--backward 
asymmetry we can determine not only the magnitude, but also the sign of the
new Wilson coefficients. 

The other experimentally measurable quantities, $\Gamma_L/\Gamma_T$ and
$\Gamma_+/\Gamma_-$, have also been studied. It is found that
$\Gamma_+/\Gamma_-$ and $\Gamma_L/\Gamma_T$ are sensitive to the $C_{RL}$
for the $B \rar K^\ast \mu^+ \mu^-$ decay. 
On the other hand, for the $B \rar K^\ast \tau^+ \tau^-$ decay, 
$\Gamma_+/\Gamma_-$ 
is more strongly dependent on $C_{RL}$ as in the $B \rar K^\ast \mu^+ \mu^-$
case, while $\Gamma_L/\Gamma_T$ is more sensitive to the coefficients 
$C_{LRLR}$ or $C_{LRRL}$. 
As the final concluding remark, we state that, from the combined analyses of 
partially integrated differential branching ratio, lepton forward--backward 
asymmetry and 
ratios of $\Gamma_+/\Gamma_-$ and $\Gamma_L/\Gamma_T$ for the 
$B \rar K^\ast \mu^+ \mu^-$ and $B \rar K^\ast \tau^+ \tau^-$ decays,
we can unequivocally  determine the existence of new physics beyond the
Standard Model, and in particular 
we can obtain  information about the various new Wilson coefficients. 
\\

%\newpage

\section*{Acknowledgements}

We sincerely thank G. Cvetic and M. Savc{\i} for useful discussions. 
The work of C.S.K. was supported 
in part by  grant No. 1999-2-111-002-5 from the Interdisciplinary 
Research Program of the KOSEF,
in part by the BSRI Program, Ministry of Education, Project No. 98-015-D00061,
in part by KRF Non-Directed-Research-Fund, Project No. 1997-001-D00111,
and in part by the KOSEF-DFG large collaboration project, 
Project No. 96-0702-01-01-2.
The work of Y.G.K. was supported by KOSEF Postdoctoral Program.

\newpage

\section*{Figure captions}

{\bf Fig. 1.} Differential branching ratio, $d {\cal BR}$ / $dq^2$ 
for $B \rightarrow K^* \mu^+ \mu^-$. 
The thick solid lines indicates standard model case, {\it i.e,} $C_X = 0$.
The thin solid, dashed, dotted and dot-dashed line correspond to
$ C_X = -C_{10}, -0.7 \times C_{10}, 0.7 \times C_{10}, C_{10} $ cases, respectively.
Here (a) $C_X = C_{LL}$ and (b) $C_X = C_{LR}$. \\ \\
{\bf Fig. 2.} Same as Fig. 1. 
Here (a) $C_X = C_{RR}$ and (b) $C_X = C_{RL}$. \\ \\
{\bf Fig. 3.} Same as Fig. 1.
Here (a) $C_X = C_{LRLR}$ and (b) $C_X = C_{LRRL}$. \\ \\
{\bf Fig. 4.} Differential branching ratio, $d {\cal BR}$ / $dq^2$ 
for $B \rightarrow K^* \tau^+ \tau^-$. 
The thick solid lines indicates standard model case, {\it i.e,} $C_X = 0$.
The thin solid, dashed, dotted and dot-dashed line correspond to
$ C_X = -C_{10}, -0.7 \times C_{10}, 0.7 \times C_{10}, C_{10} $ cases, respectively.
Here (a) $C_X = C_{LL}$ and (b) $C_X = C_{LR}$.\\ \\
{\bf Fig. 5.} Same as Fig. 4. 
Here (a) $C_X = C_{RR}$ and (b) $C_X = C_{RL}$.\\ \\
{\bf Fig. 6.} Same as Fig. 4.
Here (a) $C_X = C_{LRLR}$ and (b) $C_X = C_{LRRL}$.\\ \\
{\bf Fig. 7.} The dependence of the partially integrated 
branching ratio on the new Wilson coeffecients. 
The range for the integration is chosen 
(a) $1$ GeV$^2 < q^2 < 8$ GeV$^2$ 
for the $B \rightarrow K^* \mu^+ \mu^-$ decay and
(b) $15$ GeV$^2 < q^2 < 20$ GeV$^2$ 
for the $B \rightarrow K^* \tau^+ \tau^-$ decay.
The thick solid, thin solid, thick dashed, thin dashed, dotted and
dot-dashed line correspond to $C_X = C_{LL}, C_{LR}, C_{RL}, C_{RR},
C_{LRLR}$ and $C_{LRRL}$, respectively.\\ \\
{\bf Fig. 8.} Differential forward-backward asymmetry, $d A_{FB}$ / $dq^2$ 
for $B \rightarrow K^* \mu^+ \mu^-$. 
The thick solid lines indicates standard model case, {\it i.e,} $C_X = 0$.
The thin solid, dashed, dotted and dot-dashed line correspond to
$ C_X = -C_{10}, -0.7 \times C_{10}, 0.7 \times C_{10}, C_{10} $ cases, respectively.
Here (a) $C_X = C_{LL}$ and (b) $C_X = C_{LR}$.\\ \\
{\bf Fig. 9.} Same as Fig. 8.
Here (a) $C_X = C_{RR}$ and (b) $C_X = C_{RL}$.\\ \\
{\bf Fig. 10.} Same as Fig. 8.
Here (a) $C_X = C_{LRLR}$ and (b) $C_X = C_{LRRL}$.\\ \\
{\bf Fig. 11.} Differential forward-backward asymmetry, $d A_{FB}$ / $dq^2$ 
for $B \rightarrow K^* \tau^+ \tau^-$. 
The thick solid lines indicates standard model case, {\it i.e,} $C_X = 0$.
The thin solid, dashed, dotted and dot-dashed line correspond to
$ C_X = -C_{10}, -0.7 \times C_{10}, 0.7 \times C_{10}, C_{10} $ cases, respectively.
Here (a) $C_X = C_{LL}$ and (b) $C_X = C_{LR}$.\\ \\
{\bf Fig. 12.} Same as Fig. 11.
Here (a) $C_X = C_{RR}$ and (b) $C_X = C_{RL}$.\\ \\
{\bf Fig. 13.} Same as Fig. 11.
Here (a) $C_X = C_{LRLR}$ and (b) $C_X = C_{LRRL}$.\\ \\
{\bf Fig. 14.} The dependence of  (a) $\Gamma_+$ / $\Gamma_-$ and 
(b) $\Gamma_L$ / $\Gamma_T$ on the new Wilson coeffecients 
for $B \rightarrow K^* \mu^+ \mu^-$ decay.
The thick solid, thin solid, thick dashed, thin dashed, dotted and
dot-dashed line correspond to $C_X = C_{LL}, C_{LR}, C_{RL}, C_{RR},
C_{LRLR}$ and $C_{LRRL}$ cases.\\ \\
{\bf Fig. 15.} The dependence of  (a) $\Gamma_+$ / $\Gamma_-$ and 
(b) $\Gamma_L$ / $\Gamma_T$ on the new Wilson coeffecients 
for $B \rightarrow K^* \tau^+ \tau^-$ decay.
The thick solid, thin solid, thick dashed, thin dashed, dotted and
dot-dashed line correspond to $C_X = C_{LL}, C_{LR}, C_{RL}, C_{RR},
C_{LRLR}$ and $C_{LRRL}$ cases.\\ \\
{\bf Fig. 16.} The dependence of $\tau$ polarization on the new Wilson 
coeffecients $C_X$ for $B \rightarrow K^* \tau^+ \tau^-$ decay.
The thick solid, thin solid, thick dashed, thin dashed, dotted and
dot-dashed line correspond to $C_X = C_{LL}, C_{LR}, C_{RL}, C_{RR},
C_{LRLR}$ and $C_{LRRL}$ cases.\\ \\

\newpage

\begin{figure}[ht]
\hspace*{-0.8 truein}
\psfig{figure=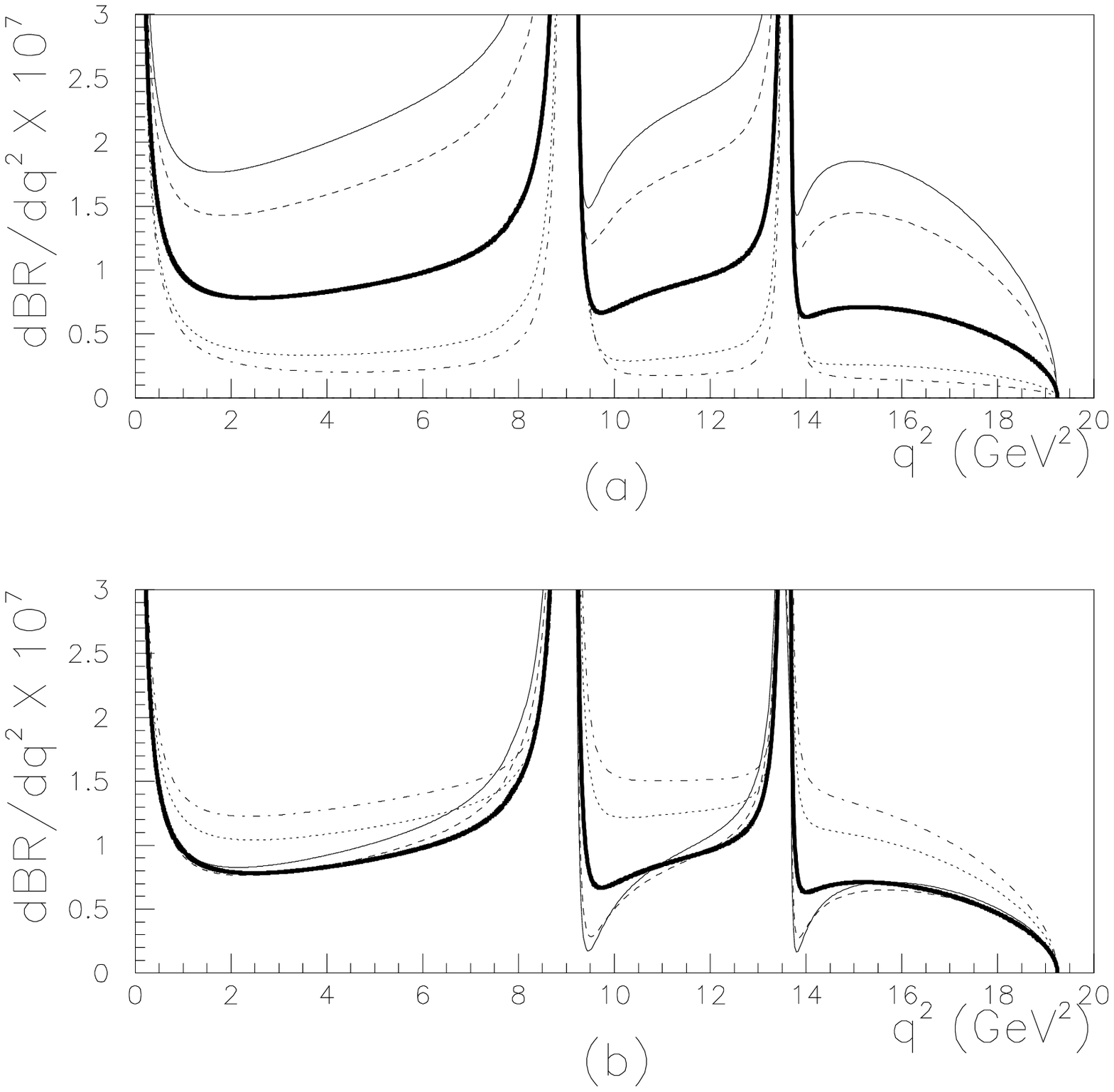}
\caption{}
\label{brm1}
\end{figure}

\begin{figure}[ht]
\hspace*{-0.8 truein}
\psfig{figure=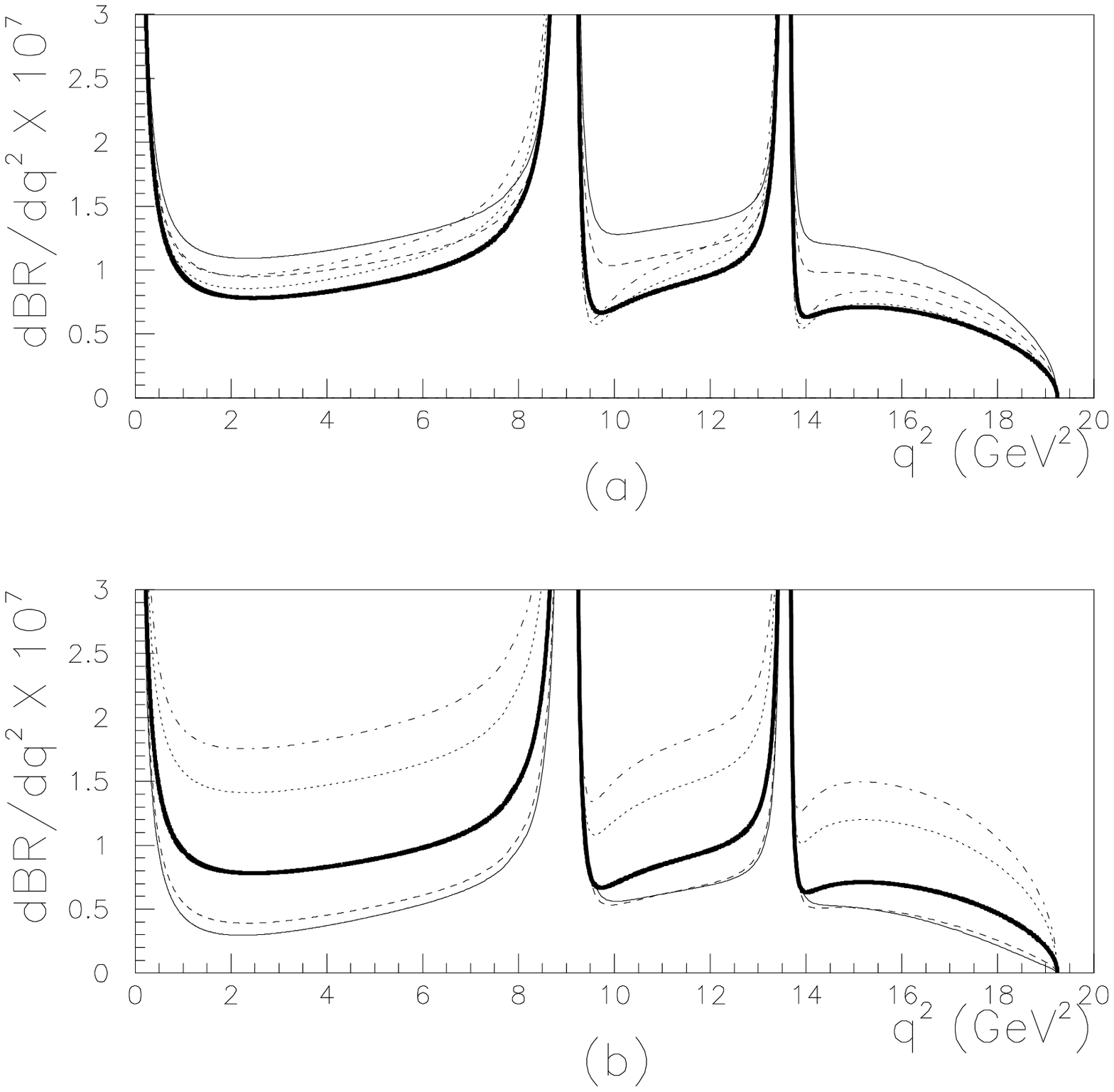}
\caption{}
\label{brm2}
\end{figure}

\begin{figure}[ht]
\hspace*{-0.8 truein}
\psfig{figure=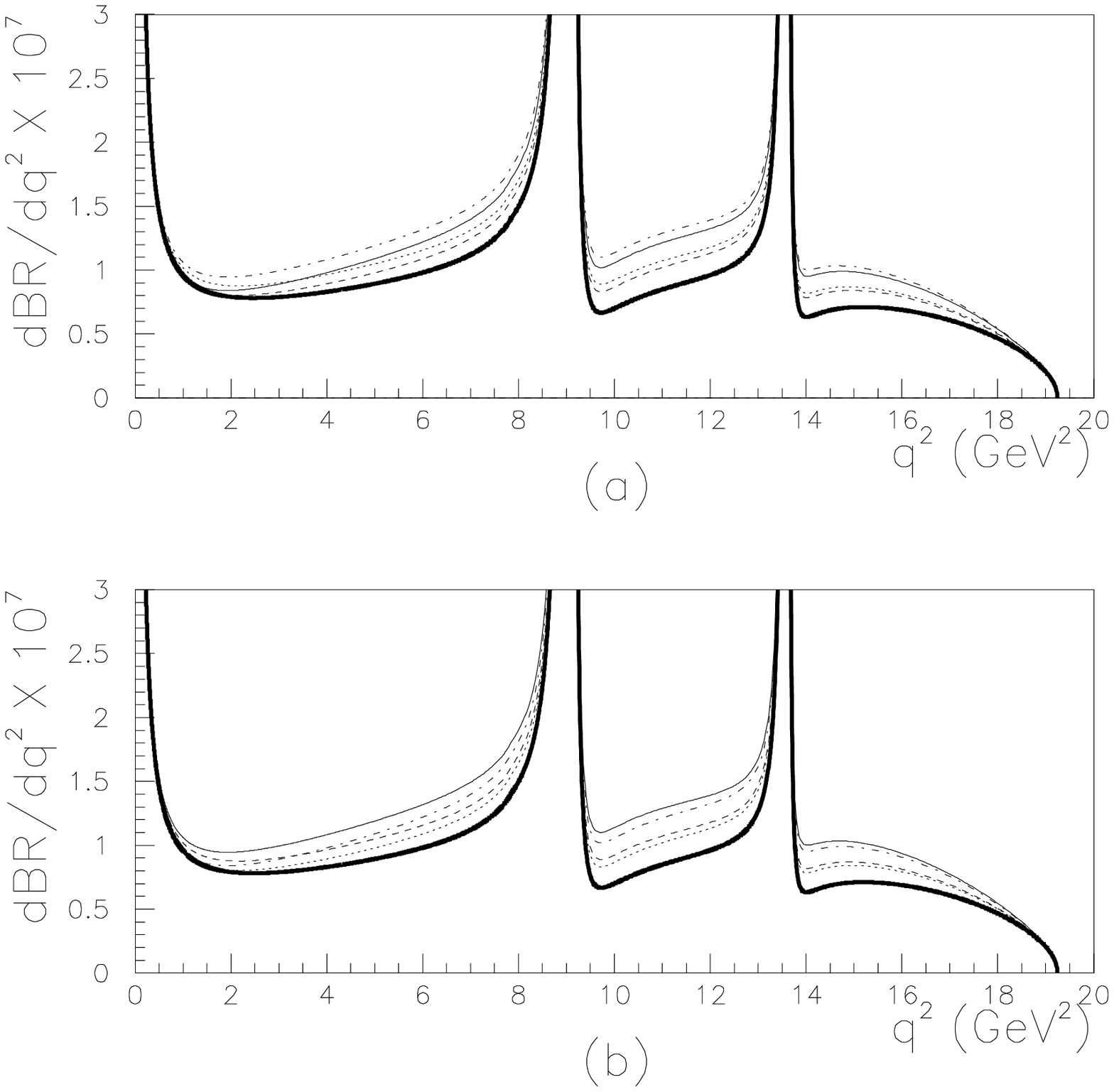}
\caption{}
\label{brm3}
\end{figure}

\begin{figure}[ht]
\hspace*{-0.7 truein}
\psfig{figure=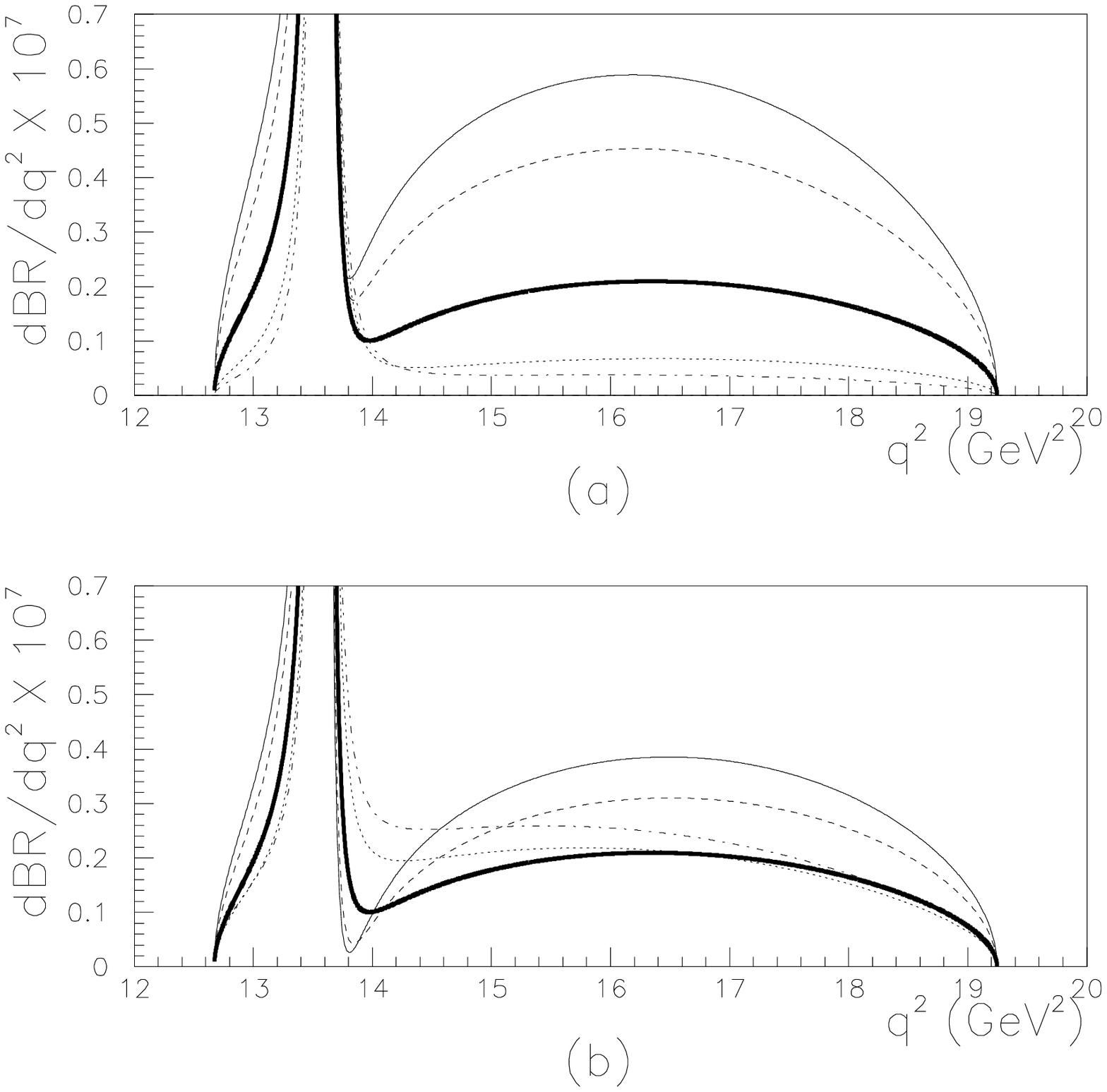}
\caption{}
\label{brt1}
\end{figure}

\begin{figure}[ht]
\hspace*{-0.7 truein}
\psfig{figure=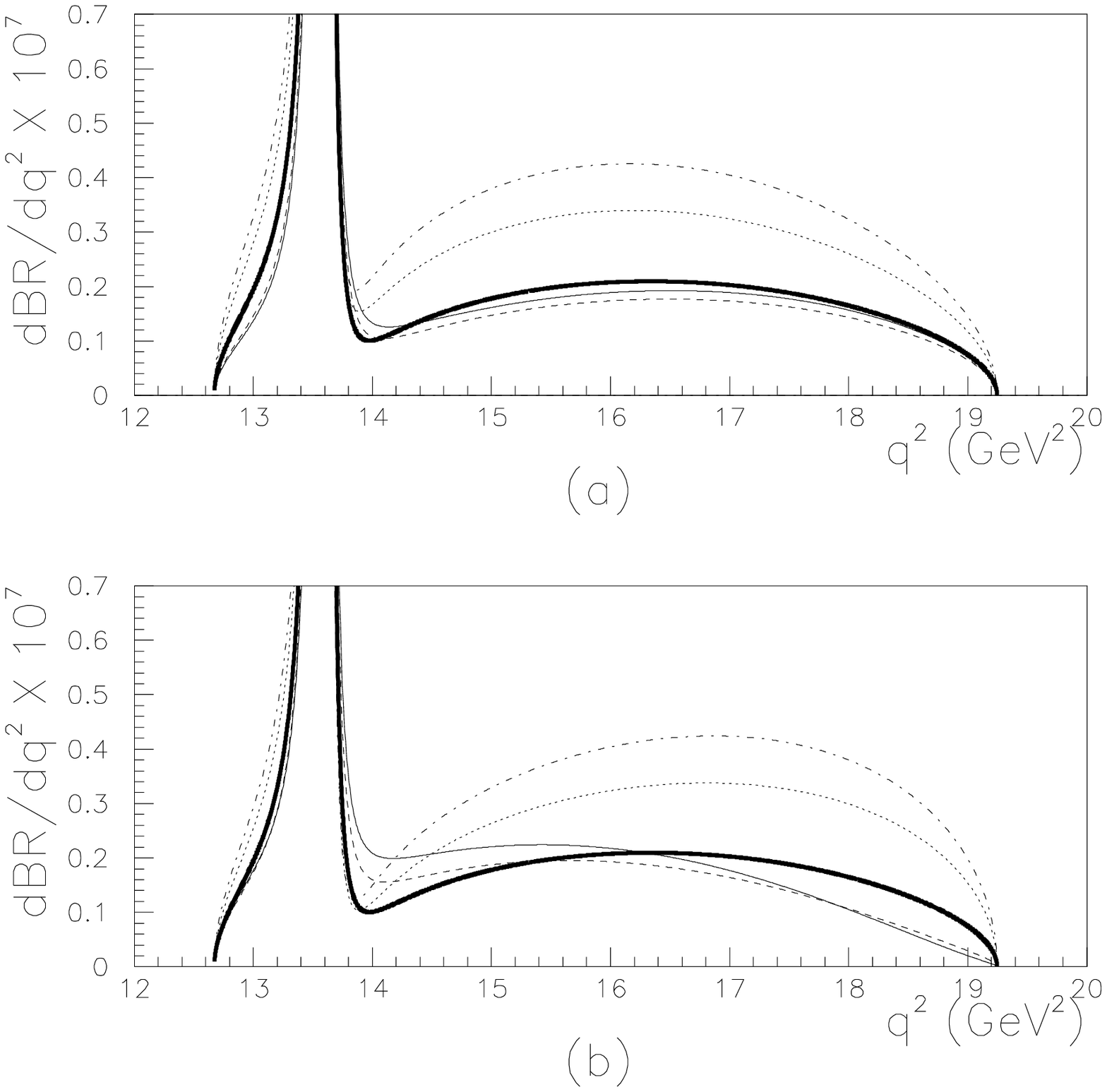}
\caption{}
\label{brt2}
\end{figure}

\begin{figure}[ht]
\hspace*{-0.7 truein}
\psfig{figure=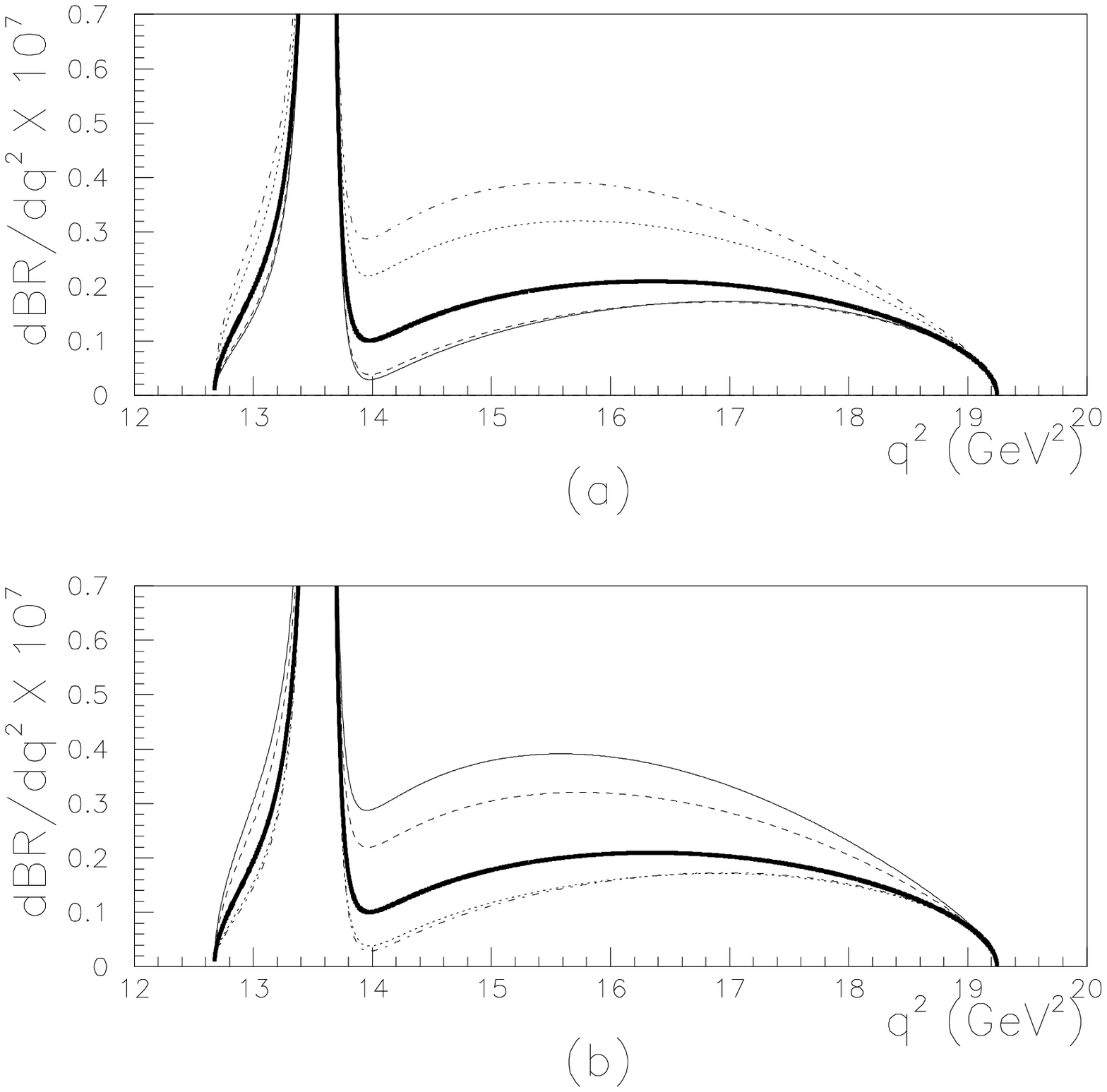}
\caption{}
\label{brt3}
\end{figure}

\begin{figure}[ht]
\hspace*{-0.7 truein}
\psfig{figure=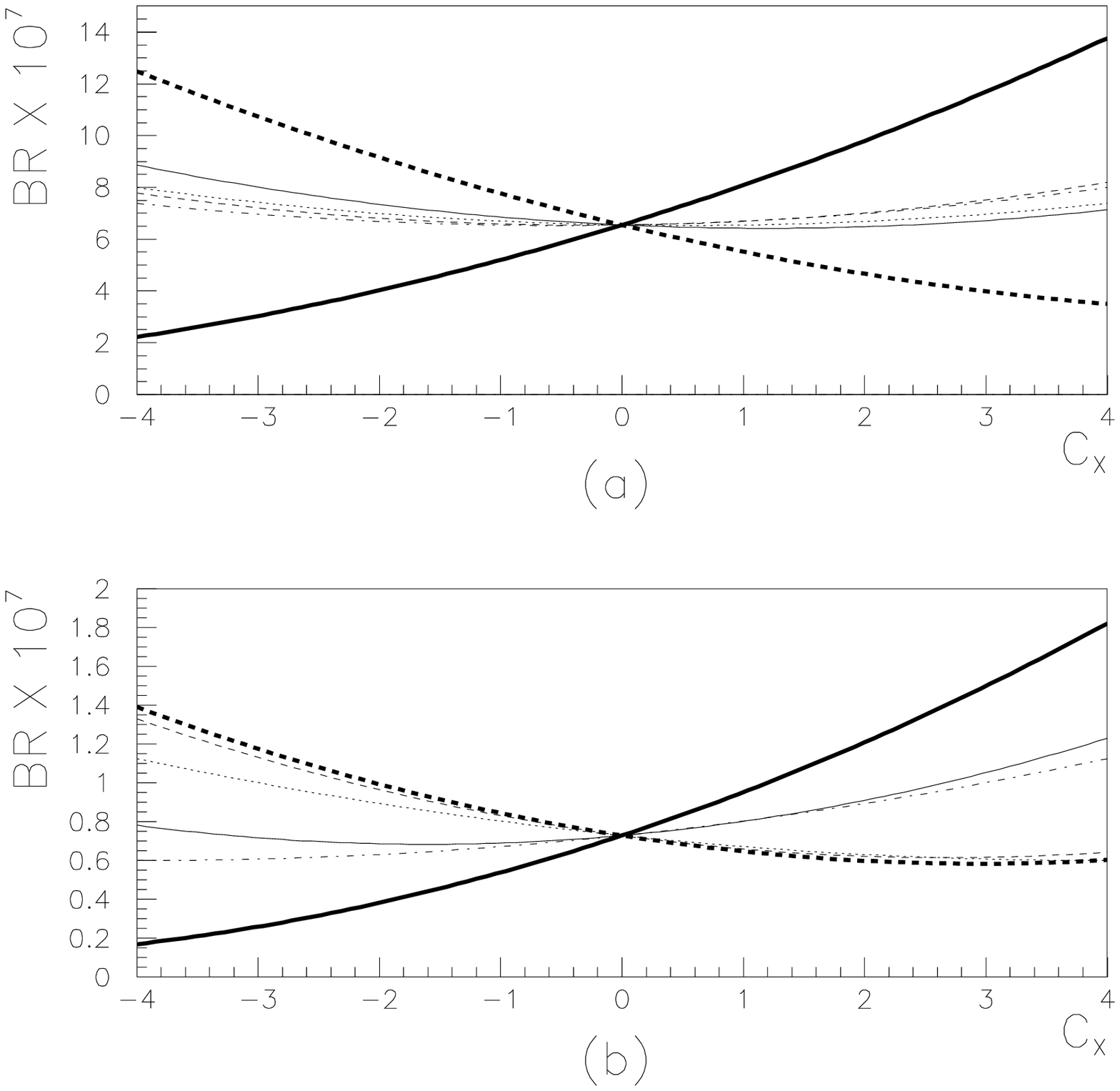}
\caption{}
\label{br}
\end{figure}

\begin{figure}[ht]
\hspace*{-0.7 truein}
\psfig{figure=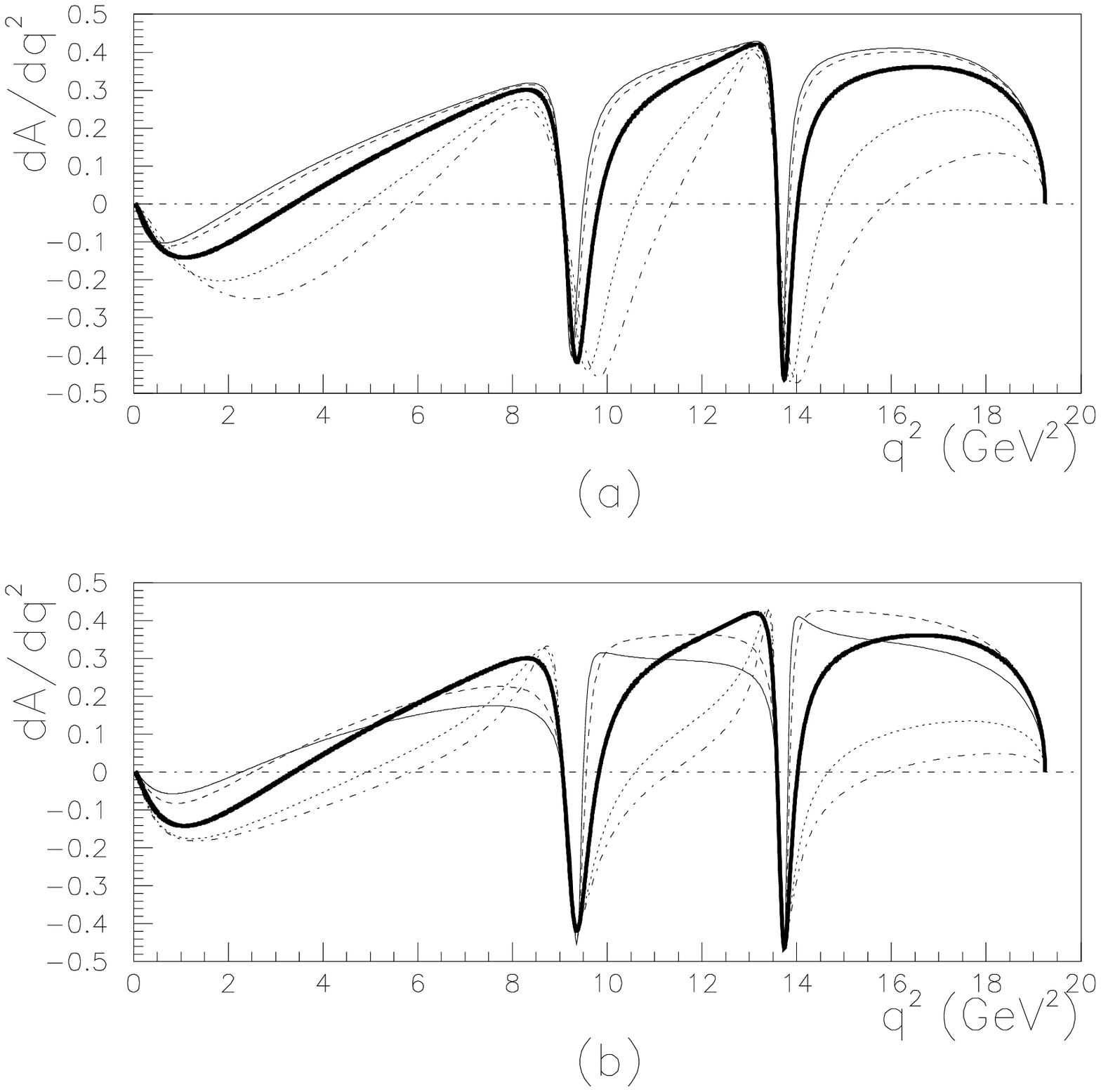}
\caption{}
\label{afbm1}
\end{figure}

\begin{figure}[ht]
\hspace*{-0.7 truein}
\psfig{figure=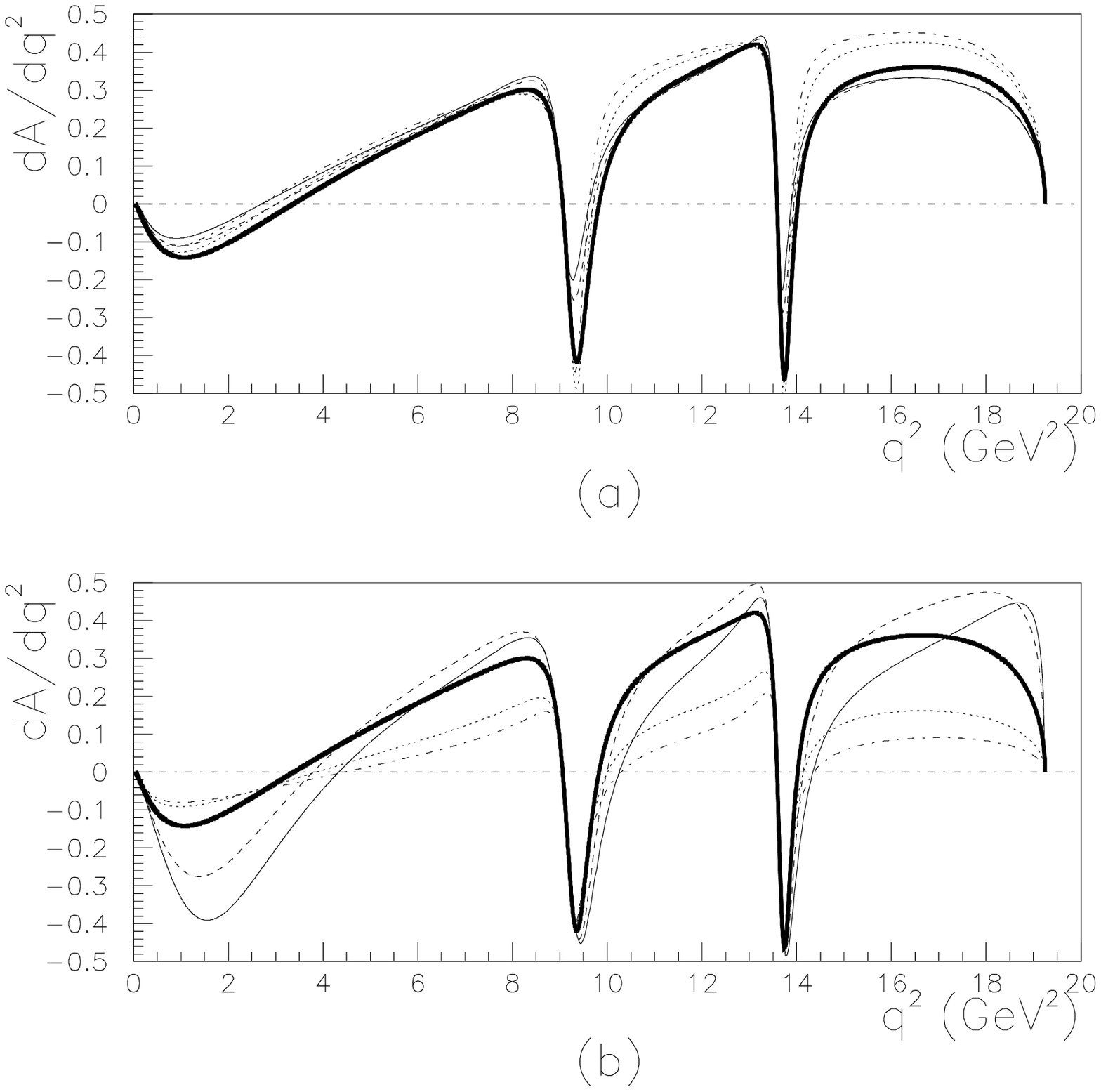}
\caption{}
\label{afbm2}
\end{figure}

\begin{figure}[ht]
\hspace*{-0.7 truein}
\psfig{figure=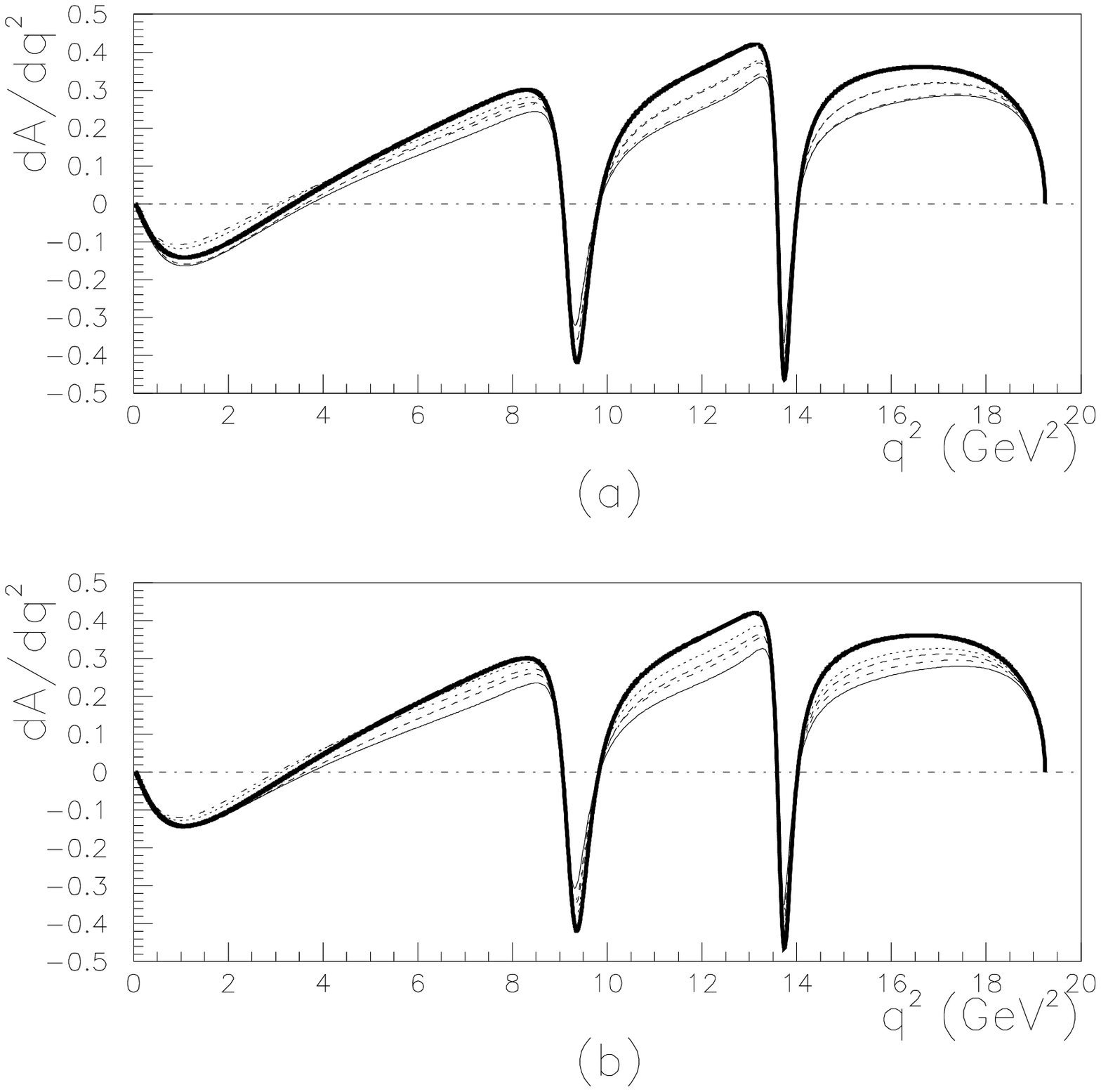}
\caption{}
\label{afbm3}
\end{figure}

\begin{figure}[ht]
\hspace*{-0.7 truein}
\psfig{figure=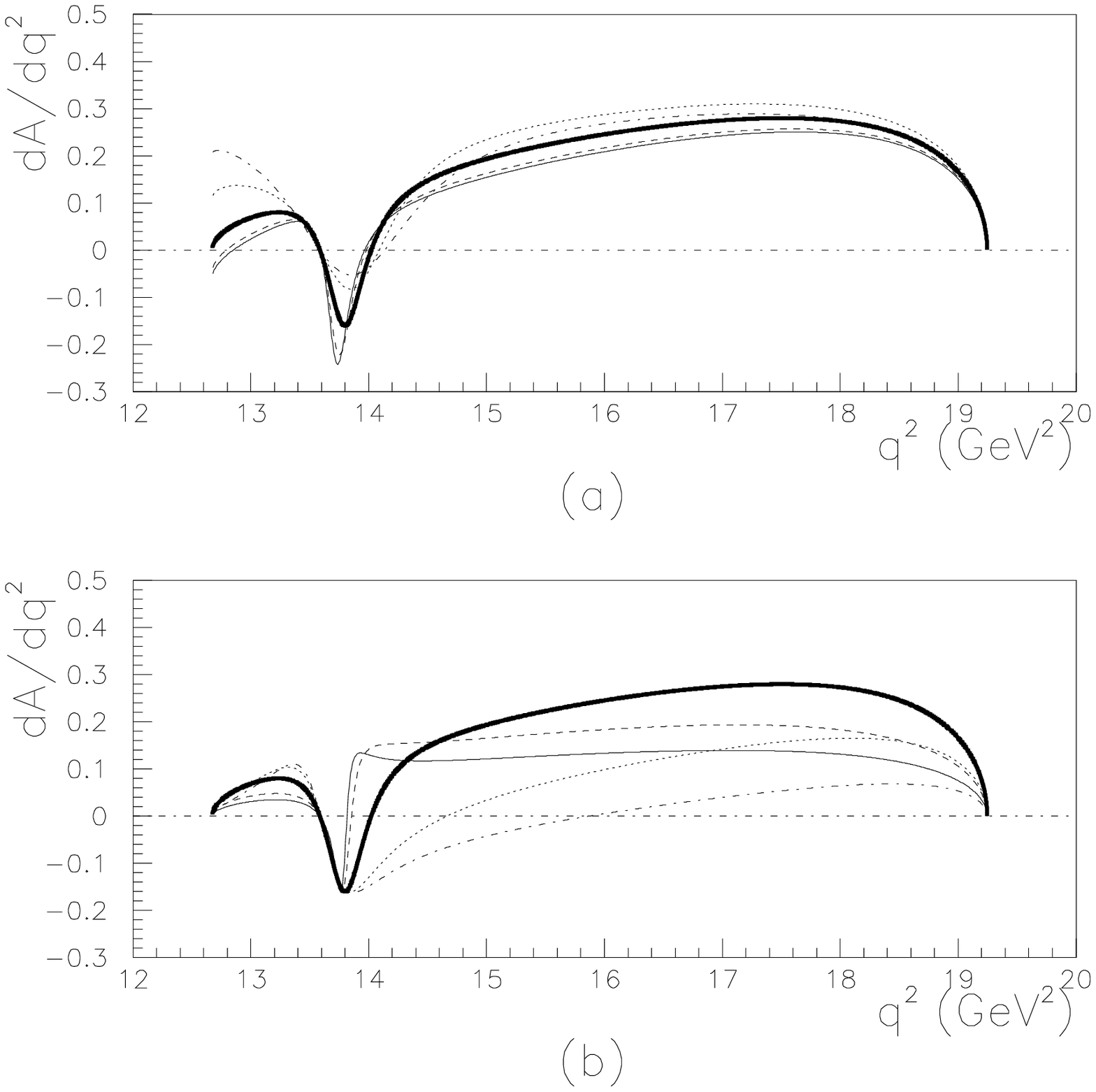}
\caption{}
\label{afbt1}
\end{figure}

\begin{figure}[ht]
\hspace*{-0.7 truein}
\psfig{figure=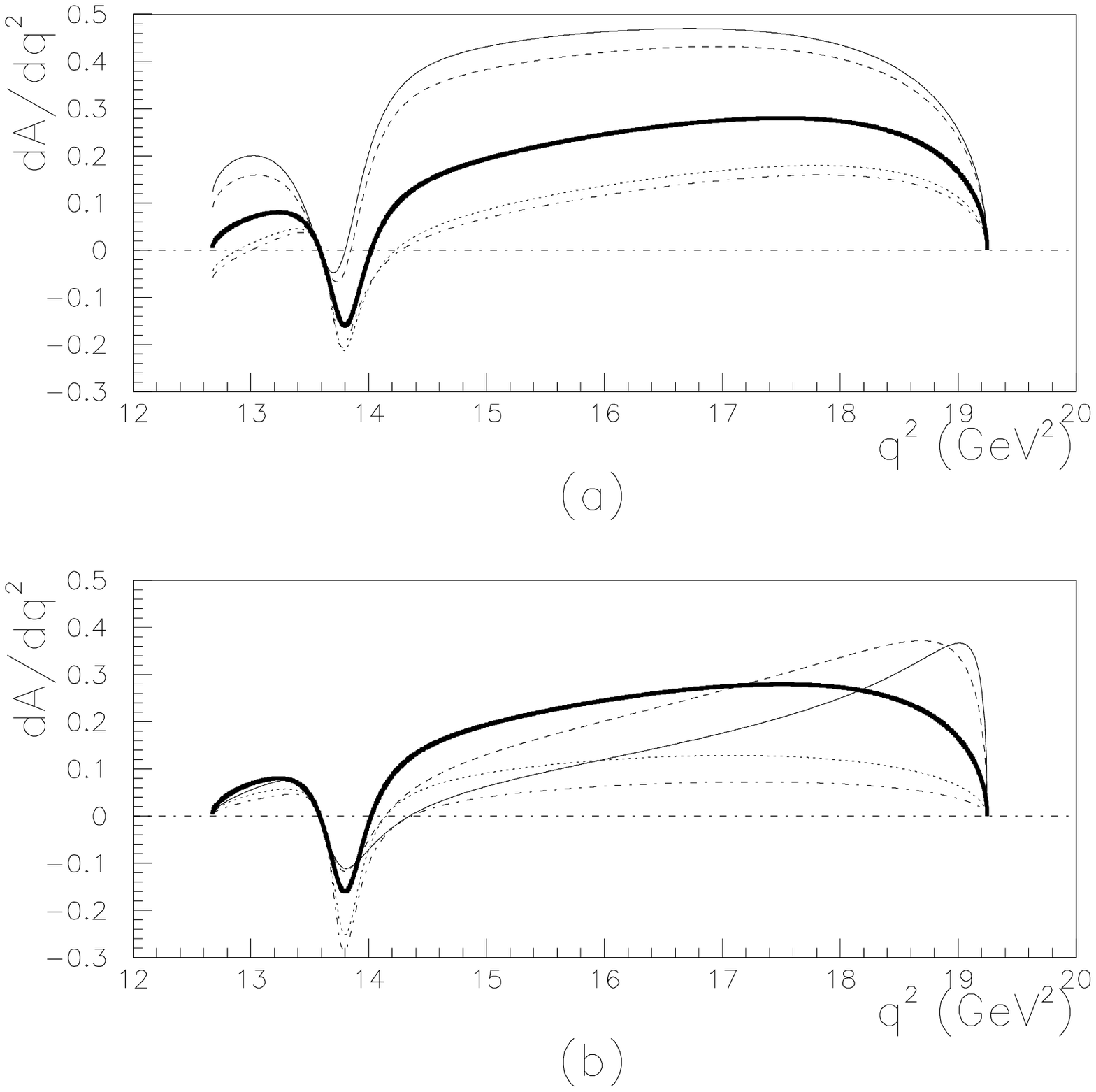}
\caption{}
\label{afbt2}
\end{figure}

\begin{figure}[ht]
\hspace*{-0.7 truein}
\psfig{figure=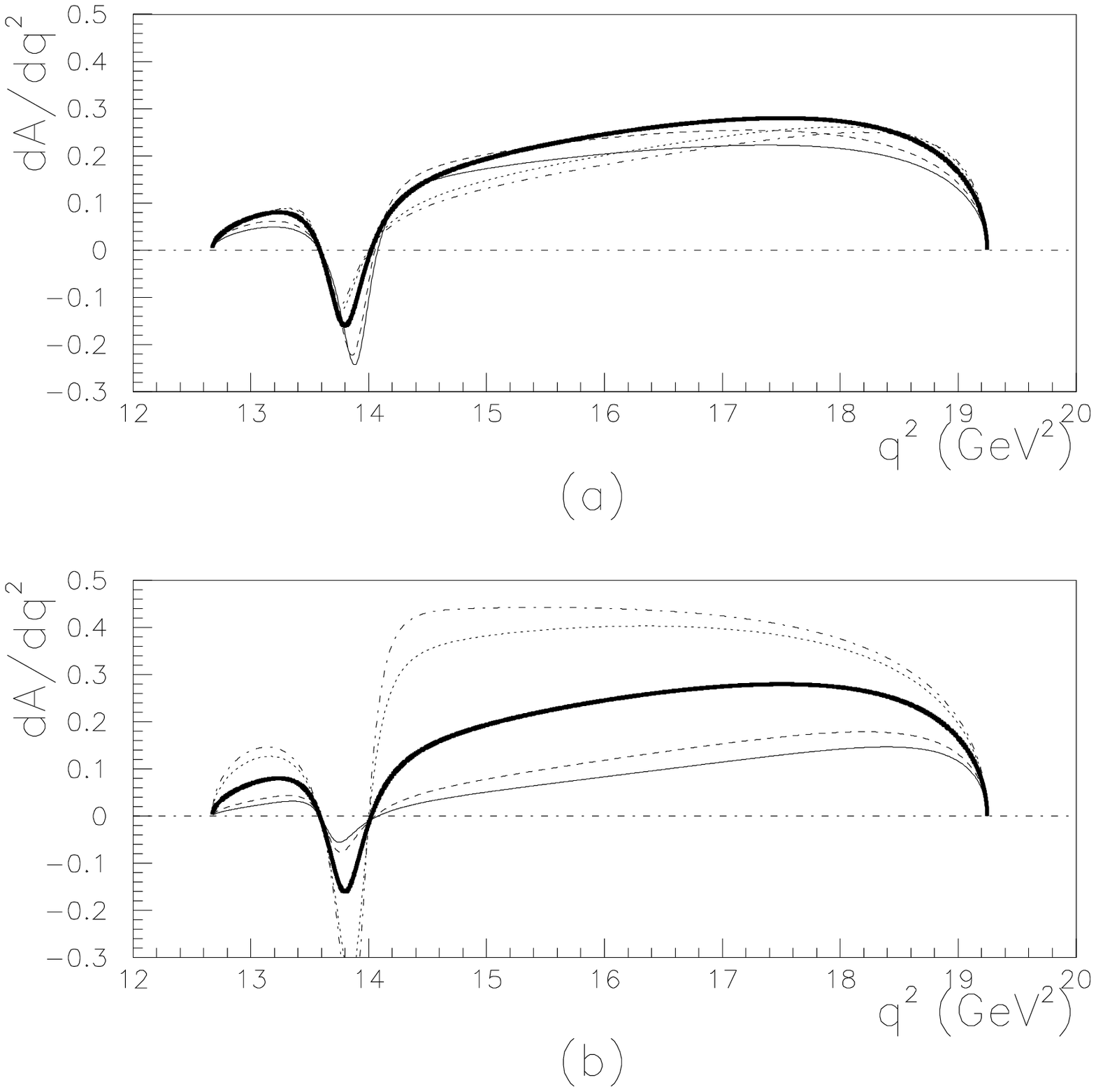}
\caption{}
\label{afbt3}
\end{figure}

\begin{figure}[ht]
\hspace*{-0.7 truein}
\psfig{figure=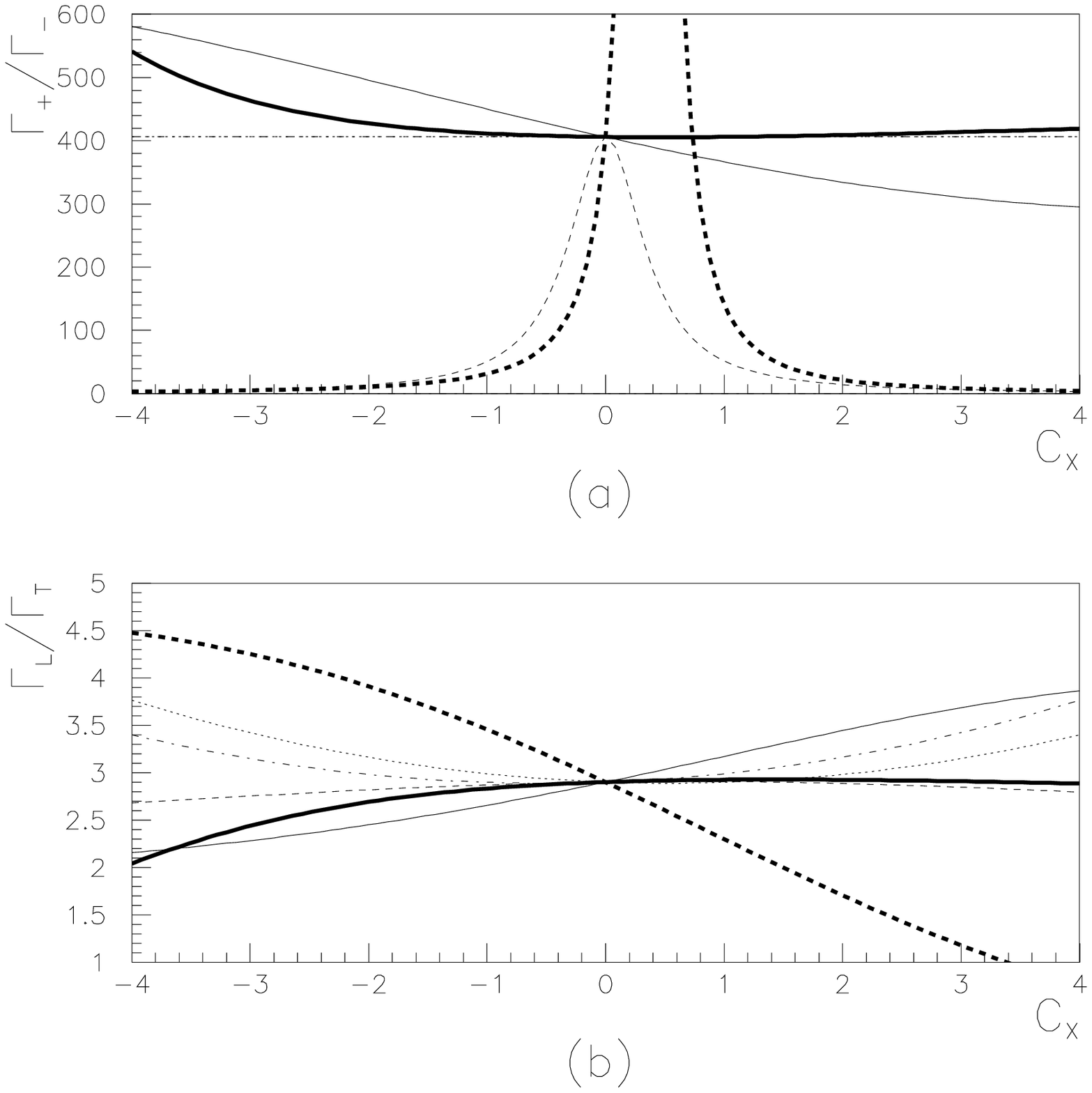}
\caption{}
\label{polm}
\end{figure}

\begin{figure}[ht]
\hspace*{-0.7 truein}
\psfig{figure=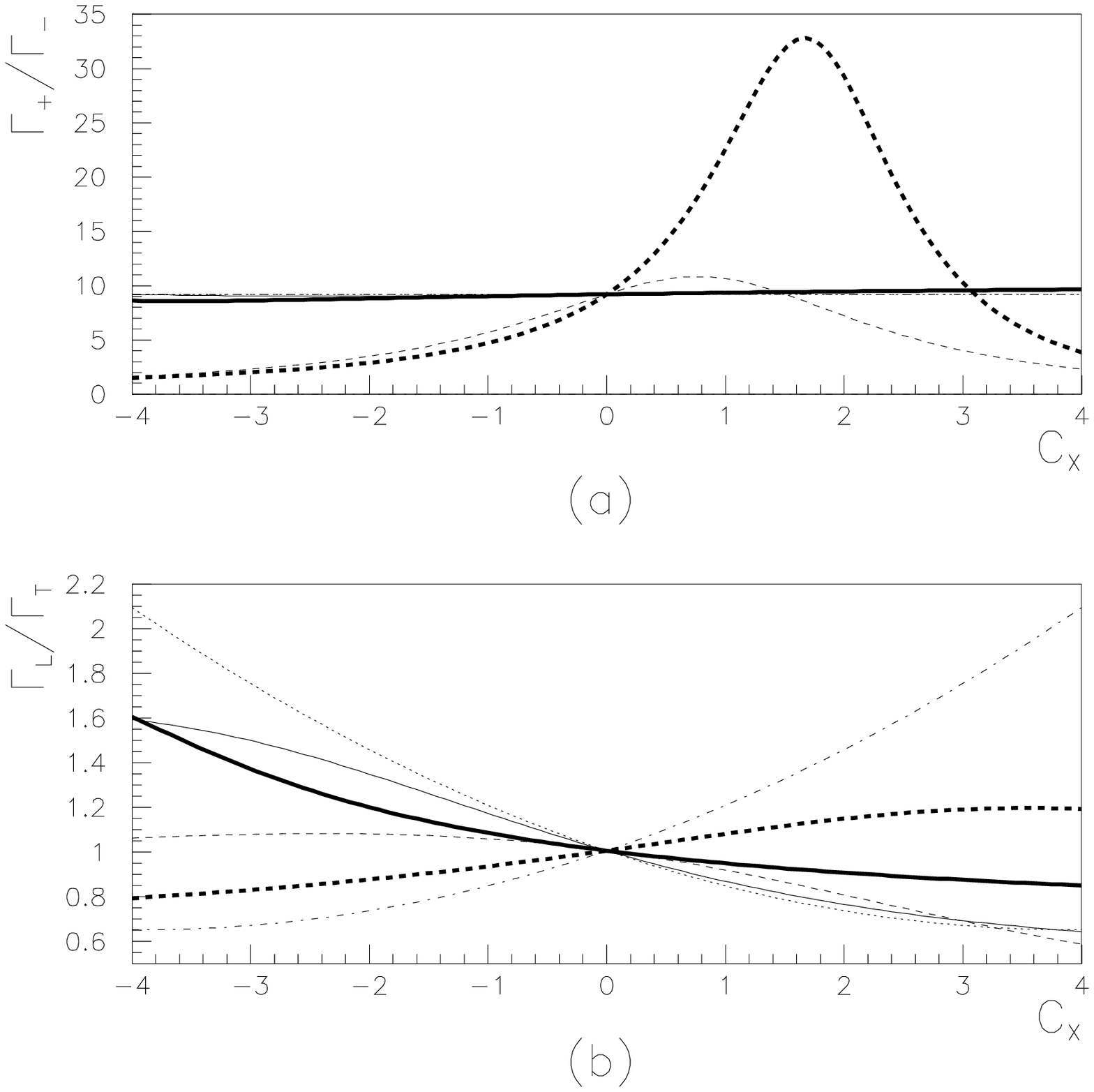}
\caption{}
\label{polt}
\end{figure}

\begin{figure}[ht]
\hspace*{-0.7 truein}
\psfig{figure=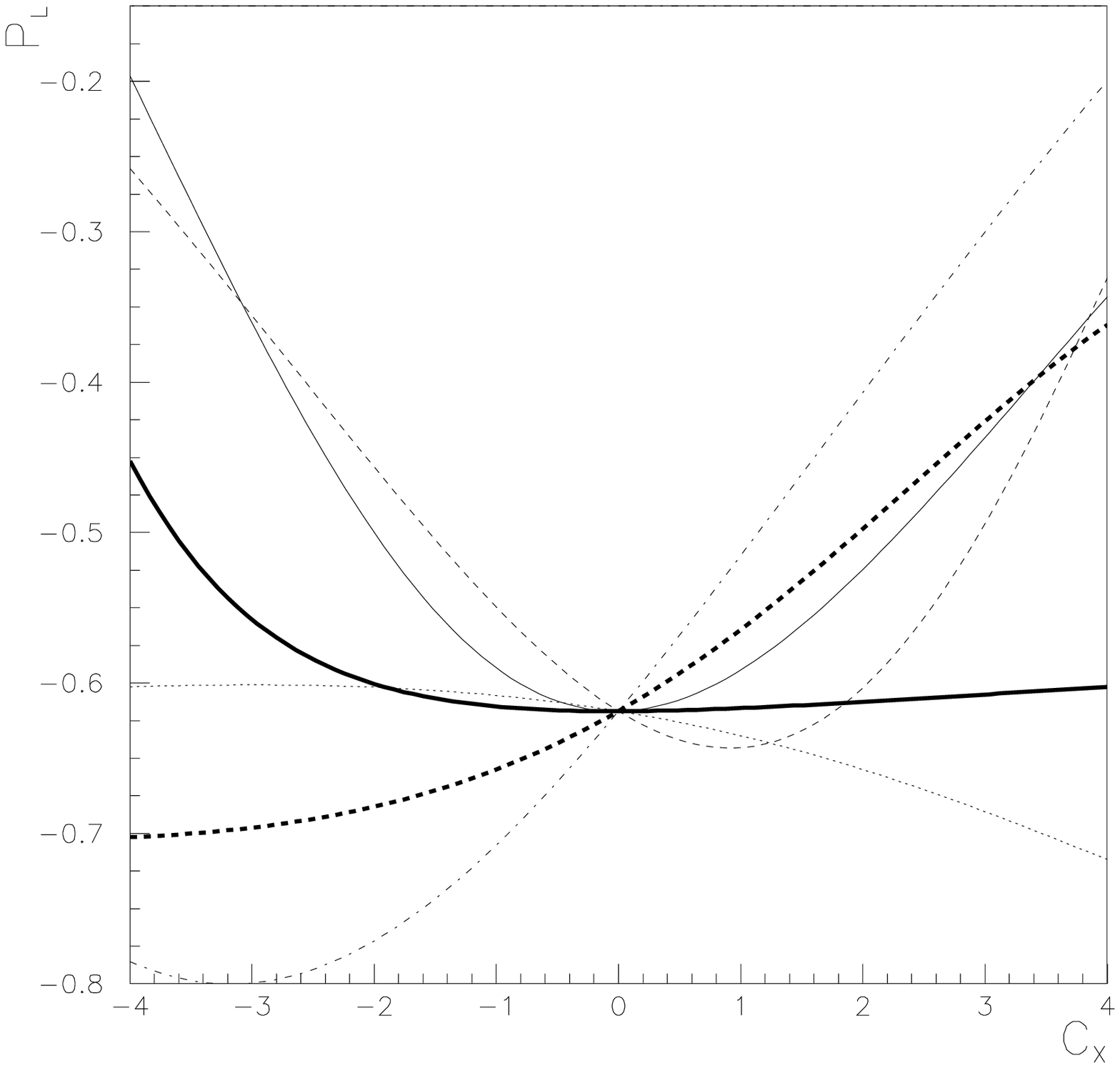}
\caption{}
\label{polt3}
\end{figure}

\end{document}